\documentclass[11pt,a4paper]{article}
\def\nl{\nonumber\\}

\def\nln{\nonumber\\*[-1ex]\phantom{\fbox{\rule{0em}{2ex}}}}

\def\beq{\begin{equation}}
\def\eeq{\end{equation}}
\def\beqar{\begin{eqnarray}}
\def\eeqar{\end{eqnarray}}
\def\barr#1{\begin{array}{#1}}
\def\earr{\end{array}}
\def\bfi{\begin{figure}}
\def\efi{\end{figure}}
\def\btab{\begin{table}}
\def\etab{\end{table}}
\def\bce{\begin{center}}
\def\ece{\end{center}}

\def\text{\textstyle}

\def\de{\delta}

\def\refeq#1{\mbox{(\ref{#1})}}

\def\refse#1{\mbox{Sect.~\ref{#1}}}
\def\refses#1{\mbox{Sects.~\ref{#1}}}

\def\refapp#1{\mbox{Appendix~\ref{#1}}}
\def\citere#1{\mbox{Ref.~\cite{#1}}}
\def\citeres#1{\mbox{Refs.~\cite{#1}}}

\def\solid{\raise.9mm\hbox{\protect\rule{1.1cm}{.2mm}}}
\def\dash{\raise.9mm\hbox{\protect\rule{2mm}{.2mm}}\hspace*{1mm}}

\newcommand{\TeV}{\unskip\,\mathrm{TeV}}

\def\mathswitchr#1{\relax\ifmmode{\mathrm{#1}}\else$\mathrm{#1}$\fi}

\newcommand{\PW}{\mathswitchr W}
\newcommand{\PZ}{\mathswitchr Z}

\newcommand{\PH}{\mathswitchr H}

\newcommand{\Pt}{\mathswitchr t}

\def\mathswitch#1{\relax\ifmmode#1\else$#1$\fi}

\newcommand{\MW}{\mathswitch {M_\PW}}

\newcommand{\MZ}{\mathswitch {M_\PZ}}
\newcommand{\MH}{\mathswitch {M_\PH}}

\newcommand{\Mt}{\mathswitch {m_\Pt}}

\def\ie{i.e.\ }

\newcommand{\ord}{{\cal O}}

\newcommand{\U}{\mathrm{U}}

\newcommand{\rT}{{\mathrm{T}}}

\newcommand{\ri}{\mathrm{i}}

\newcommand{\ieps}{\ri\epsilon}
\newcommand{\rd}{{\mathrm{d}}}

\newcommand{\Mathematica}{{\sc Mathematica}}

\makeatletter
\newcount\@tempcntc
\def\@citex[#1]#2{\if@filesw\immediate\write\@auxout{\string\citation{#2}}\fi
  \@tempcnta\z@\@tempcntb\m@ne\def\@citea{}\@cite{\@for\@citeb:=#2\do
    {\@ifundefined
       {b@\@citeb}{\@citeo\@tempcntb\m@ne\@citea
        \def\@citea{,\penalty\@m\ }{\bf ?}\@warning
       {Citation `\@citeb' on page \thepage \space undefined}}%
    {\setbox\z@\hbox{\global\@tempcntc0\csname
b@\@citeb\endcsname\relax}%
     \ifnum\@tempcntc=\z@ \@citeo\@tempcntb\m@ne
       \@citea\def\@citea{,\penalty\@m}
       \hbox{\csname b@\@citeb\endcsname}%
     \else
      \advance\@tempcntb\@ne
      \ifnum\@tempcntb=\@tempcntc
      \else\advance\@tempcntb\m@ne\@citeo
      \@tempcnta\@tempcntc\@tempcntb\@tempcntc\fi\fi}}\@citeo}{#1}}

\def\@citeo{\ifnum\@tempcnta>\@tempcntb\else\@citea
  \def\@citea{,\penalty\@m}%
  \ifnum\@tempcnta=\@tempcntb\the\@tempcnta\else
   {\advance\@tempcnta\@ne\ifnum\@tempcnta=\@tempcntb \else
\def\@citea{--}\fi
    \advance\@tempcnta\m@ne\the\@tempcnta\@citea\the\@tempcntb}\fi\fi}
\makeatother

\let\log\ln

\newcommand{\logarsymbol}[1]{\mathrm{L}_{#1}}

\newcommand{\logarithm}[2]{\logarsymbol{#1}\left(#2\right)}

\newcommand{\deriv}{\mathrm{D}}

\newcommand{\vx}{x}
\newcommand{\vy}{y}
\newcommand{\vz}{z}
\newcommand{\vi}{i}
\newcommand{\vj}{j}
\newcommand{\vk}{k}
\newcommand{\vl}{l}
\newcommand{\vm}{m}
\newcommand{\vn}{n}

\newcommand{\vp}{p}
\newcommand{\va}{a}
\newcommand{\vb}{b}
\newcommand{\vvb}{q}
\newcommand{\vc}{c}

\newcommand{\vr}{r}

\newcommand{\vt}{t}
\newcommand{\vA}{A}

\newcommand{\vT}{T}
\newcommand{\vg}{g}
\newcommand{\vgt}{\tilde g}
\newcommand{\vGt}{\tilde{G}}
\newcommand{\vghat}{\hat{g}}
\newcommand{\vfs}{f_0}
\newcommand{\vfm}{f_1}
\newcommand{\vfst}{\tilde f_0}
\newcommand{\vfmt}{\tilde f_1}
\newcommand{\vfr}{h_1}
\newcommand{\vfn}{h_0}
\newcommand{\vfnhat}{\hat{h}_0}
\newcommand{\valpha}{\alpha}
\newcommand{\vbeta}{\beta}
\newcommand{\vgamma}{\gamma}
\newcommand{\vtau}{\tau}

\newcommand{\project}{R}

\newcommand{\fun}{\vfn}
\newcommand{\tensor}{\gamma}
\newcommand{\combin}{F}
\newcommand{\NLL}{\stackrel{\mathrm{NLL}}{=}}
\newcommand{\LL}{\stackrel{\mathrm{LL}}{=}}

\newcommand{\esp}{e}
\newcommand{\espt}{{\tilde e}}

\newcommand{\numerator}{\mathcal{N}}
\newcommand{\numeratort}{\tilde\mathcal{N}}

\newcommand{\intdelta}{\bar\delta}

\newcommand{\product}[3]{P\!\left(\vec{#2}_{#3}^{\,#1}\right)}
\newcommand{\mass}[3]{\rd^{#1} \vec{#2}_{#3}}

\newcommand{\Alett}[2]{B^{#2}}

\newcommand{\ratio}{w}

\def\draftdate{\relax}
\def\mpar#1{\relax}
\def\mua{\relax}
\def\mda{\relax}
\def\mla{\relax}
\def\draft{
\def\thtystars{******************************}
\def\sixtystars{\thtystars\thtystars}
\typeout{}
\typeout{\sixtystars**}
\typeout{* Draft mode!
         For final version remove \protect\draft\space in source file *}
\typeout{\sixtystars**}
\typeout{}
\def\draftdate{\today}
\def\mua{\marginpar[\boldmath\hfil$\uparrow$]%
                   {\boldmath$\uparrow$\hfil}%
                    \typeout{marginpar: $\uparrow$}\ignorespaces}
\def\mda{\marginpar[\boldmath\hfil$\downarrow$]%
                   {\boldmath$\downarrow$\hfil}%
                    \typeout{marginpar: $\downarrow$}\ignorespaces}
\def\mla{\marginpar[\boldmath\hfil$\rightarrow$]%
                   {\boldmath$\leftarrow $\hfil}%
                    \typeout{marginpar: $\leftrightarrow$}\ignorespaces}
\def\Mua{\marginpar[\boldmath\hfil$\Uparrow$]%
                   {\boldmath$\Uparrow$\hfil}%
                    \typeout{marginpar: $\Uparrow$}\ignorespaces}
\def\Mda{\marginpar[\boldmath\hfil$\Downarrow$]%
                   {\boldmath$\Downarrow$\hfil}%
                    \typeout{marginpar: $\Downarrow$}\ignorespaces}
\def\Mla{\marginpar[\boldmath\hfil$\Rightarrow$]%
                   {\boldmath$\Leftarrow $\hfil}%
                    \typeout{marginpar: $\Leftrightarrow$}\ignorespaces}
\def\mpar##1{\marginpar{\hbadness10000%
                      \sloppy\hfuzz10pt\boldmath\bf##1}%
                      \typeout{marginpar: ##1}\ignorespaces}

\overfullrule 5pt
\oddsidemargin -15mm
\marginparwidth 29mm
}
\newcommand{\thismonth}{\ifcase\month\or January\or February\or March \or April
\or May \or June \or July \or August \or September \or \November \or 
\December\fi}

\makeatletter

\def\eqnarray{\stepcounter{equation}\let\@currentlabel=\theequation
\global\@eqnswtrue
\global\@eqcnt\z@\tabskip\@centering\let\\=\@eqncr
$$\halign to \displaywidth\bgroup\hskip\@centering
  $\displaystyle\tabskip\z@{##}$\@eqnsel&\global\@eqcnt\@ne
  \hskip 2\arraycolsep \hfil${##}$\hfil
  &\global\@eqcnt\tw@ \hskip 2\arraycolsep $\displaystyle\tabskip\z@{##}$\hfil
   \tabskip\@centering&\llap{##}\tabskip\z@\cr}
\makeatother

\begin{document}

\thispagestyle{empty} 

\thispagestyle{empty}
\def\thefootnote{\fnsymbol{footnote}}
\setcounter{footnote}{1}
\null
\draftdate\hfill SFB/CPP-04-31\\
\strut\hfill   TTP04-17\\
\strut\hfill PSI-PR-04-09\\
\strut\hfill hep-ph/0408068
\vskip 0cm
\vfill
\begin{center}
{\Large \bf
An algorithm for the high-energy expansion of multi-loop diagrams to next-to-leading logarithmic accuracy
\par}
 \vskip 2.5em
{\large
{\sc A.~Denner\footnote{ansgar.denner@psi.ch} }}
\\[1ex]
{\it Paul Scherrer Institut \\
CH-5232 Villigen PSI, Switzerland
}\\[2ex]
{\large
{\sc S.~Pozzorini\footnote{pozzorin@particle.uni-karlsruhe.de} }}
\\[1ex]
{\it Institut f\"ur Theoretische Teilchenphysik, 
Universit\"at Karlsruhe \\
D-76128 Karlsruhe, Germany}
\par
\end{center}\par
\vskip 1.0cm 
\vfill 

{\bf Abstract:} \par 

We present an algorithm to compute arbitrary multi-loop massive
Feynman diagrams in the region where the typical energy scale
$\sqrt{s}$ is much larger than the typical mass scale $M$, \ie $s\gg
M^2$, while various different energy and mass parameters may be
present.  In this region we perform an asymptotic expansion and, using
sector decomposition, we extract the leading contributions
resulting from ultraviolet and mass singularities, which consist of
large logarithms $\log(s/M^2)$ and $1/\varepsilon$ poles in
$D=4-2\varepsilon$ dimensions.  To next-to-leading accuracy, at $L$
loops all terms of the form $\alpha^L \varepsilon^{-k} \log^j(s/M^2)$
with $j+k=2L$ and $j+k=2L-1$ are taken into account.  This algorithm
permits, in particular, to compute higher-order next-to-leading
logarithmic electroweak corrections for processes involving various
kinematical invariants of the order of hundreds of GeV and masses
$\MW\sim\MZ\sim\MH\sim\Mt$ of the order of the electroweak scale, in
the approximation where the masses of the light fermions are
neglected.


\par
\vskip 1cm
\noindent
August 2004 
\par
\null
\setcounter{page}{0}
\clearpage
\def\thefootnote{\arabic{footnote}}
\setcounter{footnote}{0}
\newpage

\section{Introduction}
With the advent of colliders in the TeV energy range, multi-loop
calculations, both in QCD and in the electroweak theory, will play an
increasingly important role in order to provide theoretical
predictions of sufficiently high accuracy.  On the one hand, this is
due to the high level of experimental precision, especially at the
linear collider \cite{Aguilar-Saavedra:2001rg,Abe:2001wn,Abe:2001gc}.
On the other hand, the size of radiative corrections grows with the
energy.

This latter feature is particularly important for the electroweak
corrections \cite{Kuroda:1991wn}, since at TeV colliders the energy
scale $\sqrt{s}$ starts to be much larger than the weak-boson mass
scale $M=\MW\sim\MZ$.  For $M^2\ll s$, \ie if one performs an
asymptotic expansion in $M^2/s$, the electroweak corrections assume
the form of a tower of large logarithms,
\beq\label{logtower}
\alpha^L \sum_{j=0}^{2L}a_j 
\log^{j}\left(\frac{s}{M^2}\right),
\eeq
at $L$-loop level. These logarithms originate from ultraviolet (UV)
and mass singularities. Above the electroweak scale, such logarithms
start to grow and become the leading contribution to the electroweak
corrections.  At $\sqrt{s}=1\TeV$ they give rise to corrections of
tens of per cent at one loop and a few per cent at two loops.  Thus,
in order to reach the per-mille level accuracy envisaged by the linear
collider, it becomes mandatory to compute electroweak radiative
corrections up to the two-loop level.

Despite of recent progress in this direction \cite{Aglietti:2003yc},
the exact analytical computation of two-loop electroweak corrections
for processes involving many (more than three) external legs and
various internal masses is out of sight.  One is therefore forced
either to adopt a numerical approach, as for instance in
\citeres{Passarino:2001wv,Binoth:2000ps}, or to consider an asymptotic
expansion of the type \refeq{logtower}, as we do in the present paper.
 
The asymptotic behaviour of higher-order electroweak corrections has
received a lot of interest in the recent years
\cite{Kuhn:1999de,Ciafaloni:2000ub,Fadin:2000bq,Kuhn:2000nn,Melles:2001gw,%
Kuhn:2001hz,Ciafaloni:2000df,Melles:2000ed,Hori:2000tm,Beenakker:2000kb,%
Denner:2003wi,Pozzorini:2004rm,Feucht:2003yx,Feucht:2004rp}.  On the
one hand, resummation prescriptions have been proposed
\cite{Kuhn:1999de,Ciafaloni:2000ub,Fadin:2000bq,Kuhn:2000nn,Melles:2001gw},
which predict the leading logarithms (LLs) and next-to-leading
logarithms (NLLs), \ie terms with $j=2L$ and $2L-1$ in
\refeq{logtower}, for arbitrary processes and also the
next-to-next-to-leading logarithms (NNLLs), with $j=2L-2$, for $2\to
2$ massless fermionic processes \cite{Kuhn:2001hz}.  On the other
hand, in order to check these resummations, which rely on the
assumption that various aspects of the symmetry-breaking mechanism can
be neglected in the high-energy limit, explicit diagrammatic two-loop
calculations have been performed.  For the LLs
\cite{Melles:2000ed,Hori:2000tm,Beenakker:2000kb} as well as for the
angular-dependent subset of the NLLs \cite{Denner:2003wi} such checks
have been already completed and the exponentiation predicted in
\citeres{Fadin:2000bq,Kuhn:2000nn,Melles:2001gw} has been confirmed
for arbitrary processes.  At (or beyond) the level of the NLLs only
very few calculations exist.  In the electroweak Standard Model, a
calculation of a massless fermionic form factor
\cite{Pozzorini:2004rm} confirmed the resummation prescriptions of
\citeres{Kuhn:2000nn,Melles:2001gw} up to the NLLs.  The evaluation of
the complete tower of logarithms contributing to this form factor in
the special case of a $\U(1)\times\U(1)$ theory with a massive and a
massless gauge boson \cite{Feucht:2003yx,Feucht:2004rp} was found to
be in agreement with the resummation prescriptions of
\citere{Kuhn:2001hz} at the NNLL level.

For general processes, the terms with $j=2L-1$ in the asymptotic
expansion \refeq{logtower} remain largely untested and those for $j\le
2L-2$ almost completely unknown.  It is important to further
investigate the behaviour of these subleading logarithms since,
depending on the process, their phenomenological impact at the TeV
scale might be comparable with that of the LLs, as was found in
\citeres{Kuhn:2000nn,Kuhn:2001hz,Feucht:2003yx,Feucht:2004rp} for
processes involving massless fermions.

In this paper we present an algorithm to compute the LLs and NLLs in a
fully automated way.  This algorithm, which has already been used in
\citeres{Denner:2003wi,Pozzorini:2004rm}, applies to arbitrary
multi-loop Feynman diagrams involving various energy parameters
$s_j\sim s$ and masses $M_k\sim M$ of the same order.  In the
asymptotic region $s\gg M^2$, the coefficients of the LLs and NLLs are
computed analytically as a function of the ratios $s_j/s$ and $M_k/M$.
The mass singularities originating from soft/collinear massless
particles (photons, gluons, or fermions) as well as the UV
singularities are regulated dimensionally and appear as
$1/\varepsilon$ poles in $D=4-2\varepsilon$ dimensions.

The algorithm is based on the so-called sector decomposition, a
technique that permits to separate overlapping UV or mass
singularities in Feynman-parameter (FP) integrals.  Originally, this
method was introduced in order to deal with UV singularities
\cite{Hepp:1966eg}. Later, it turned out to be applicable to mass
singularities \cite{Roth:thesis,Roth:1996pd}, and a first general
algorithm to isolate $1/\varepsilon$ poles in massless loop integrals
has been formulated in \citere{Binoth:2000ps}.  The algorithm that we
present here permits to extract combinations (products) of
$1/\varepsilon$ poles and mass-singular logarithms of $s/M^2$, which
arise in the case where both massless and massive particles are
present.

The input is an arbitrary multi-loop tensor integral, which is first
translated into a FP integral.  Depending on the location of UV and
mass singularities in FP space, the integration range is decomposed
into sectors, and each sector is remapped into the original
integration range. This can be done iteratively until, in each sector,
all singularities arise when individual FPs tend to zero and these FPs
can be factorized.  As a consequence, a simple power counting permits
to determine the degree of singularity, \ie the power of logarithms
and $1/\varepsilon$ poles resulting from each sector.  One can thus
easily select those sectors that are responsible for the leading and
next-to-leading singularities.  Furthermore, those FP integrations
that lead to singularities have a standard structure and can be
computed in analytic form.

The coefficients of the resulting logarithms and $1/\varepsilon$ poles
are represented as integrals over the remaining FPs.  As a result of
sector decomposition, these last integrations are free of
singularities and, within the Euclidean region, the integrands are
smooth functions which can be integrated numerically.  Moreover, the
coefficients of the leading and next-to-leading singularities are very
simple integrals and in all examples we considered up to now (two-loop
diagrams with up to 4 external legs and various internal masses) they
could be solved in analytic form by standard computer algebra programs
such as \Mathematica.
 
The paper is organized as follows: in \refse{se:NLLA} we describe the
basics of the high-energy limit and the logarithmic approximation. In
\refse{se:Feynmanintegrals} we review the general expressions for
multi-loop FP integrals and their UV and mass singularities and sketch
the idea of the sector decomposition. The sector decomposition and the
extraction of the singularities of massless diagrams is presented in
\refse{se:masslessdecomp} and the treatment of massive diagrams in
\refse{se:massivedecomp}. The applicability of the proposed method is
discussed in \refse{se:discussion}. Some notations and conventions and
several results for relevant massless and massive integrals are listed
in the appendices.

\section{High-energy limit and logarithmic approximation}
\label{se:NLLA}

The computation of higher-order electroweak corrections is a
multi-scale problem.  Typically there are various energy parameters
$s_1,s_2,\dots$, which correspond to the kinematical invariants of the
process, masses of the order of the electroweak scale
$M=\MW\sim\MZ\sim\MH\sim\Mt$, various light-fermion masses $m_f\ll M$
and a fictitious infinitesimal photon mass to regulate infrared
singularities.  If all light fermions (including b quarks) and photons
are treated as massless, then all particle masses are either zero or
of the order $M$.  Assuming this approach, in the present paper we
consider the kinematical regime where all energy parameters are much
higher than the electroweak scale.  To be general, we consider
multi-loop Feynman diagrams characterized by the hierarchy
\beq\label{scales}
s \sim |s_1|\sim |s_2|\sim \dots \sim |s_J| 
\gg
M^2 \sim M^2_1\sim M^2_2\sim \dots \sim M^2_K
\eeq
of energy parameters and masses, \ie we assume that there are only two
different scales $s>0$ and $M^2>0$.  The external particles can be
either on-shell with masses of the order $M$ or zero, or off-shell
with invariant masses of the order $s$.  In this regime, we perform an
asymptotic expansion in the small mass-to-energy ratio
\beq
\ratio=\frac{M^2}{s}\ll 1,
\eeq
treating the ratios $s_j/s$ and $M_k/M$ as constants in the limit
$\ratio\to 0$.  The leading terms of this expansion are divergent.  On
the one hand there are mass singularities associated to massive
particles which appear as logarithms of $\ratio$.  On the other hand,
mass singularities from massless particles and UV singularities appear
as $1/\varepsilon$ poles in $D=4-2\varepsilon$ dimensions.  The
leading terms of the asymptotic expansion of $L$-loop Feynman
integrals\footnote{We assume that overall factors with mass dimension
  have been factorized such that the amplitude $A(\ratio)$ is
  dimensionless.}  assume the form of a double series in $\varepsilon$
and $\log(\ratio)$,
\beqar\label{tower}
A(\ratio)=\ratio^{d}
\sum_{j=0}^{2L}\sum_{k=-j}^\infty 
a_{j,k}\, \varepsilon^k  \log^{j+k}(\ratio) + \ord(\ratio^{d+1}).
\eeqar
In general, the loop integrals can give rise to non-logarithmic mass
singularities, \ie $d$ can be negative.  However, in spontaneously
broken gauge theories such singularities are compensated by mass terms
or coupling constants proportional to masses in such a way that the
resulting contributions are at most logarithmically divergent.  In the
present paper, we compute loop integrals restricting ourselves to the
leading terms of order $\ratio^d$ whereas contributions of order
$\ratio^{d+1}$, which are suppressed by powers of masses, are
neglected.

The various terms in \refeq{tower} are classified according to their
degree of singularity $j$, which corresponds to the total power of
logarithms and $1/\varepsilon$ poles.  For a fixed value of $j$ we
compute all terms of the order
\beq\label{polestimeslogs}
\varepsilon^{-j},
\varepsilon^{-j+1}\log(\ratio),
\varepsilon^{-j+2}\log^2(\ratio),
\dots,
\log^{j}(\ratio),
\varepsilon\log^{j+1}(\ratio),
\dots,
\eeq
up to the needed order in $\varepsilon$.  At $L$-loop level, all terms
of the type \refeq{polestimeslogs} with $j=2L, 2L-1,2L-2,\dots$, are
collectively denoted as leading logarithms (LLs), next-to-leading
logarithms (NLLs), next-to-next-to-leading logarithms (NNLLs), and so
on.  For instance, with this convention, the logarithms of order
$\log^{2L}(\ratio)$ as well as the poles of order $\varepsilon^{-2L}$
are denoted as LLs.  In the present paper we adopt a NLL
approximation, \ie we include all contribution of the type
\refeq{polestimeslogs} with $j\ge 2L-1$.  To denote this approximation
we use the symbol $\NLL$.

Since there are different energies and masses, the coefficients of the
expansion \refeq{tower} depend on their ratios, $s_j/s$ and $M_k/M$.
This dependence is taken into account and permits, for instance, to
compute electroweak corrections without assuming that
$\MW=\MZ=\MH=\Mt$ and/or including angular-dependent corrections.

\section{Feynman-parameter integrals}
\label{se:Feynmanintegrals} 

\newcommand{\Udet}{\mathcal{U}}
\newcommand{\Udett}{\tilde\mathcal{U}}
\newcommand{\F}{\mathcal{F}}
\newcommand{\Ft}{\tilde\mathcal{F}}
In this section we review the well-known technique \cite{FPint} to
translate loop integrals into FP integrals.  Then we discuss the
origin of UV and mass singularities in FP space.

\subsection{Derivation of the Feynman-parameter representation}
\label{se:Derivation_Feynmanintegrals} 

Let us consider a generic $L$-loop Feynman integral in $D$ dimensions
with $I$ internal lines, $E$ external lines, and $m$ loop momenta in
the numerator,
\beqar\label{originalfpintegral}
G_0&=&
\int \prod_{l=1}^L 
\rd\tilde{q}_l\,
\frac{
q_{a_1}^{\mu_1}\dots
q_{a_m}^{\mu_m}
}{
\prod_{s=1}^I D_s^{n_s}
}
\;\project_{\mu_1\dots\mu_m}
,\qquad
\rd\tilde{q}_l=
\mu^{(4-D)}
\frac{\rd^D q_l}{(2\pi)^D}.
\eeqar
The propagators $D_s^{-1}=(r_s^2-m_s^2+\ieps)^{-1}$ with infinitesimal
positive imaginary part $\epsilon$, which are raised to integer powers
$n_s\ge 1$, are determined by the internal momenta $r_s$, which
consist of linear combinations of loop momenta $q=(q_1,\dots,q_L)$ and
external momenta $p=(p_1,\dots,p_E)$, depending on the topology.  The
loop momenta occurring in the numerator are contracted with a
projector $\project_{\mu_1\dots\mu_m}$, which depends in general on
the external momenta. In the following, we choose to split off a
universal factor $(4\pi\mu^2/s)^{(2-D/2)L}$ from the $L$-loop Feynman
integral in order to simplify the notation.  This is equivalent to the
choice $\mu^2=s/(4\pi)$. This factor can be trivially restored at the
end if needed.

In order to combine the products of propagators into a single 
denominator one introduces FPs by means of the well-known identity
\beqar\label{feynparamformula}
\frac{1}{
\prod_{s=1}^I D_s^{n_s}
}&=&
\frac{\Gamma(N)}{\prod_{s=1}^I\Gamma(n_s)}
\int_0^1\rd^I \vec{\alpha}\,\,
\delta\biggl(1-\sum_{s=1}^I \alpha_s\biggr)
\frac{\prod_{s=1}^I\alpha_s^{n_s-1}
}{
\left[\sum_{s=1}^I \alpha_s D_s
\right]^N
}
\eeqar
with $N=\sum_{s=1}^I n_s$.  In order to perform the integration over
the loop momenta $q$, we eliminate the linear terms from the quadratic
form in the denominator, so that
\beq
\sum_{s=1}^I\alpha_s D_s=  
\sum_{i,j=1}^{L} q_i M_{ij} q_l -2 \sum_{i=1}^L q_i Q_i - J + \ieps
=
\sum_{i,j=1}^{L} k_i M_{ij} k_l -\Delta + \ieps.
\eeq
This is easily obtained by a shift of the loop momenta with
\beq
k_l=q_l-v_l,
\qquad
v_l=\sum_{i=1}^L M^{-1}_{li} Q_i
,\qquad
\Delta= J+\sum_{i,j=1}^L Q_i M^{-1}_{ij} Q_j.
\eeq
The resulting FP integral consists of a sum of terms with different
numbers of loop momenta in the numerator.  Only terms with an even
number of loop momenta do not vanish, \ie terms of the type
\beqar\label{shiftetmomenta2} 
G&=&
\frac{\Gamma(N)}{\prod_{s=1}^I\Gamma(n_s)}
\int_0^1\rd^I\vec{\alpha}\,\delta\biggl(1-\sum_{s=1}^I \alpha_s\biggr)
\,
\prod_{s=1}^I\alpha_s^{n_s-1}\,
\nl
&&{}\times
\int \prod_{l=1}^L  \rd \tilde{k}_l\,
\frac{
k_{a_1}^{\mu_1}\cdots
k_{a_{2p}}^{\mu_{2p}}
}{
\left[k^\rT M k -\Delta +\ieps
\right]^N
}\,
v^{\mu_{2p+1}}_{a_{2p+1}}\cdots
v^{\mu_{m}}_{a_{m}}
\;\project_{\mu_1\dots\mu_m}
\eeqar
with $0\le 2p\le m$.  The integration over the loop momenta yields
\beqar
\lefteqn{
\Gamma(N)
\int \prod_{l=1}^L  \rd \tilde{k}_l\,
\frac{
k_{a_1}^{\mu_1}\dots
k_{a_{2p}}^{\mu_{2p}}
}{
\left[k^\rT M k -\Delta+\ieps
\right]^N
}=}\quad&&\nl&=&
(-1)^{N-p}
\left[\frac{\ri s^{2-D/2}
}{ (4\pi)^{2}}\right]^L
\Gamma\left(N-\frac{LD}{2}-p\right)
\frac{\left(\det{M}\right)^{-D/2}}{(\Delta-\ieps)^{N-LD/2-p}}\,
\tensor^{\mu_1\dots\mu_{2p}}_{a_1\dots a_{2p}}(M^{-1}),
\nln
\eeqar
where $\gamma(M^{-1})=1$ for $p=0$, whereas for $p\ge 1$
\beqar
\tensor^{\mu_1\dots\mu_{2p}}_{a_1\dots a_{2p}}(M^{-1})&=&
\frac{1}{2^p}\sum_{i_1\dots i_p \atop j_1\dots j_p}
\prod_{k=1}^p \left(M^{-1}\right)_{a_{i_k}a_{j_k}}
g^{\mu_{i_k}\mu_{j_k}}
\eeqar
with the sum running over all $(2p)!/(p!\, 2^p)$ pairings
$(i_1,j_1),\dots,(i_p,j_p)$ obtained from the set $\{1,2,\dots,2p\}$.
As a result we are left with a FP integral of the form
\beqar\label{fpoutput}
G&=&
\int_0^1\rd^I\vec{\alpha}\,\,\delta\biggl(1-\sum_{s=1}^I \alpha_s\biggr)
\,\vghat(\vec{\alpha}).
\eeqar
In order to isolate the inverse powers of $\det M$ which originate
from $M^{-1}$, we write the integrand $\vghat(\vec{\alpha})$ in terms
of$\,$\footnote{Note that we factorize the energy scale $s$ in order
  to define $\F$ as a dimensionless function.}%
\beq
\Udet=\det{M},\qquad
\tilde{M}^{-1}_{ij}=M^{-1}_{ij}\det{M},\qquad
\F= s^{-1}\Delta \det{M}-\ieps,\qquad
w_i= v_i\det{M},
\eeq
resulting in 
\beqar\label{fpoutput1}
\vghat(\vec \alpha)=
\Gamma(\esp+L\varepsilon)
\frac{\numerator(\vec{\alpha})}
{\left[\F(\vec{\alpha})\right]^{\esp+L\varepsilon}
\left[\Udet(\vec{\alpha})\right]^{f-(L+1)\varepsilon}}
\eeqar
in $D=4-2\varepsilon$ dimensions, with $\esp=N-2L-p$, $f=2(L+1)-N+m$
and
\beqar\label{fpoutput2}
\numerator(\vec{\alpha})&=&
(-1)^{N-p}
s^{-e}\left[\frac{\ri}{ (4\pi)^{2}}\right]^L
\prod_{s=1}^I
\frac{\alpha_s^{n_s-1}}{\Gamma(n_s)}\,
\tensor^{\mu_1\dots\mu_{2p}}_{a_1\dots a_{2p}}(\tilde{M}^{-1})\,
w^{\mu_{2p+1}}_{a_{2p+1}}\cdots
w^{\mu_{m}}_{a_{m}}
\;\project_{\mu_1\dots\mu_m}
.
\eeqar
\subsection{Ultraviolet and mass singularities}

As discussed in \refse{se:NLLA}, the large logarithmic contributions
that dominate the high-energy expansion of loop diagrams originate
from UV and mass singularities.  Let us now discuss the origin of such
singularities in the FP integrals \refeq{fpoutput}--\refeq{fpoutput2}.

The overall gamma function $\Gamma(\esp+\varepsilon)$ in
\refeq{fpoutput1} represents an obvious source of UV singularities if
$\esp\le 0$.  In addition, there are less trivial UV and mass
singularities that originate from end-point singularities of the
integrand.  Let us focus on this kind of divergences, which result
from the zeros of the polynomials $\Udet$ and/or $\F$ when their
exponents, $f$ and $e$, are positive.

The zeros of the function $\Udet$, which are independent of the
external masses and momenta, are associated to UV subdivergences.
Instead, the zeros of the function $\F$ can always be avoided by
choosing non-zero internal and external masses and thus correspond to
mass singularities. These latter, as is well known, originate from
soft and/or collinear regions in momentum space.

It is useful to recall that the functions $\Udet$ and $\F$, which are
independent of the loop momenta occurring in the numerator of
\refeq{originalfpintegral}, are determined by the topology of the
corresponding Feynman diagram \cite{FPint}.  First, $\Udet$ is an
homogeneous polynomial of degree $L$ in the FPs,
\beq\label{Udef}
\Udet(\vec{\alpha})=\sum_{\mathcal{T}}
\prod_{k=1\atop i_k \notin \mathcal{T}}^L \alpha_{i_k},
\eeq
where the sum runs over all trees $\mathcal{T}$ of the Feynman
diagram.  A tree is obtained by cutting $L$ lines ($i_1,\dots,i_L$) of
the diagram such that all vertices remain connected.  Second, $\F$ is
a homogeneous polynomial of degree $L+1$ in the FPs,
\beq\label{Fdef}
s \F(\vec{\alpha})=-\sum_{\mathcal{C}} s_{\mathcal{C}} \prod_{k=1
  \atop i_k \in \mathcal{C}}^{L+1} \alpha_{i_k} +
\Udet(\vec{\alpha})\sum_{s=1}^I \alpha_s m_s^2 - \ieps ,
\eeq
where the first sum runs over all cuts $\mathcal{C}$ of the Feynman
diagram. A cut is a set of $L+1$ lines ($i_1,\dots,i_{L+1}$) that when
cut divide the diagram in two connected subdiagrams.  Finally,
\beq\label{Fdef2}
s_{\mathcal{C}}=\left(\sum_{k=1}^{L+1} \tilde{r}_{i_k} \right)^2
\quad\mbox{with}\quad
\tilde{r}_{i_k}=\pm {r}_{i_k}
\eeq
denotes the squared total momentum%
\footnote{ The signs in \refeq{Fdef2} must be chosen in such a way
  that all momenta $\tilde{r}_{i_k}$ have the same direction with
  respect to the cut $\mathcal{C}$.}  flowing through the cut
$\mathcal{C}$.

All coefficients of the polynomial $\Udet$ are non-negative.  This
implies that the UV subdivergences, which result from the zeros of
this polynomial, originate only from regions of the type
\beq\label{singularregions}
\{\vec{\alpha}\,|\,\alpha_{j_1}=\dots=\alpha_{j_n}=0\}
\quad \mbox{with}\quad  1\le n \le I-1,
\eeq
where one or more FPs vanish. Here $\{j_1,\ldots, j_n\}$ is a subset
of $\{1,\ldots,I\}$.  Mass singularities originate from the zeros of
the polynomial $\F$ in the limit of vanishing internal and external
masses, where $\F$ depends only on the invariants $s_{\mathcal{C}}$
that are not squares of on-shell external momenta, \ie the invariants
$s_j$.  In the case where these invariants are negative ($s_j<0$) all
coefficients of the polynomial \refeq{Fdef} are non-negative for
vanishing internal and external masses, and mass singularities only
originate from regions of the type \refeq{singularregions}.  This is
valid for arbitrary, \ie also positive, values of the invariants $s_j$
since mass singularities appear independently of the directions of the
external momenta \cite{Kinoshita:1962ur} and thus of the values (and
the signs) of these invariants.

\subsection{Subtraction of overlapping singularities}
\label{se:subtraction}

Using the fact that all singularities we are interested in originate
from regions of the type \refeq{singularregions}, we can isolate them
at the level of the integrand and simplify their computation.  This
can be done with simple subtractions that split the integrand into
singular and non-singular parts in such a way that the integration of
the singular part is drastically simplified.

Let us first illustrate how to proceed in the simple case of a
one-dimensional integral assuming that the integrand has the form
$1/[\alpha \vfs(\alpha)]$ with a polynomial $\vfs$ such that
$\vfs(\alpha)>0$ for $0\le \alpha \le 1$.  Here, the logarithmic
singularity can be isolated by means of a simple subtraction at
$\alpha=0$:
\beq\label{onedimexample1}
\int_0^1 \frac{\rd \alpha}{\alpha} 
\frac{1}{{\vfs}(\alpha)}
=
\underbrace{\int_0^1 \frac{\rd \alpha}{\alpha} 
\frac{1}{{\vfs}(0)}}_{\mbox{singular}}
+
\underbrace{\int_0^1 \frac{\rd \alpha}{\alpha} 
\left[
\frac{1}{{\vfs}(\alpha)}
-
\frac{1}{{\vfs}(0)}
\right]}_{\mbox{finite}}
.
\eeq
As a result, the singular part, which requires the introduction of an
appropriate regularization prescription, becomes much simpler than the
original integral.  Also the remaining non-singular part is simplified
since the integrand is a smooth function and can be integrated, either
analytically or numerically, in the absence of a regulator.

In the case of massive integrals, the mass-to-energy ratio $w=M^2/s$
plays the role of the regulator and the mass-singular logarithms that
arise in the high-energy limit, $w\ll 1$, can be isolated by means of
a subtraction similar to \refeq{onedimexample1}.  If we assume that
the integrand has the form $1/[\alpha \vfs(\alpha)+w \vfm(\alpha)]$,
with polynomials $\vfs$ and $\vfm$ such that $f_i(\alpha)>0$ for $0\le
\alpha\le 1$, then we can write
\beq\label{onedimexample2}
\int_0^1 
\frac{\rd \alpha}{\alpha {\vfs}(\alpha)+w \vfm(\alpha)}
=
\int_0^1 
\frac{\rd \alpha}{\alpha {\vfs}(0)+w \vfm(0)}
+\int_0^1 \frac{\rd \alpha}{\alpha} 
\left[
\frac{1}{{\vfs}(\alpha)}
-
\frac{1}{{\vfs}(0)}
\right]
+\ord(w),
\eeq
and the  mass-suppressed terms of order $w$ can be  neglected.

In order to be able to perform the above subtractions, it is crucial
that the polynomial which leads to the singularity is written in the
form $\alpha \vfs(\alpha)$, where the integration variable $\alpha$,
which is responsible for the singularity at $\alpha=0$, is {\em
  factorized}.  In the above one-dimensional examples this is always
possible, since any polynomial with a zero at $\alpha=0$ can be
written in this form.  Instead, in the case of multi-dimensional FP
integrals, the situation is complicated by the presence of so-called
{\em overlapping} singularities, which originate only from regions
where different FPs vanish simultaneously, \ie regions of the type
\refeq{singularregions} with $n>1$.  The polynomials that are
responsible for these overlapping singularities can be written in the
form 
$\sum_{k=1}^n \alpha_{j_k}f_{0,k}(\vec\alpha)$ 
and the FPs
$\alpha_{j_k}$ cannot be factorized.

This difficulty can be overcome by means of the so-called
sector-decomposition technique, which consists of a decomposition of
the integration range into various sectors followed by a remapping of
each sector into the original integration range.  As we show in the
following two sections, the sectors can be chosen in such a way that
the resulting integrals are of the form
\beq\label{multidimexample1}
\int_0^1 \frac{\rd \vec \alpha}
{\Bigl(\prod_{j} \alpha_j\Bigr) {\vfs}(\vec \alpha)
+w \vfm (\vec\alpha)
},
\eeq
where all FPs which are responsible for the singularities are
factorized.

\section{Sector decomposition of massless diagrams}
\label{se:masslessdecomp} 
\newcommand{\psect}{P}
\newcommand{\fpset}{R}
The sector decomposition of the FP integrals
\refeq{fpoutput}--\refeq{fpoutput2} consists of two steps.  First a
primary sector decomposition is performed that permits to eliminate
the $\delta$-function in such a way that the regions
\refeq{singularregions}, which give rise to singularities, remain
unchanged.  In the second step, an iterated sector decomposition that
permits to factorize all FPs that lead to singularities as discussed
in the previous section is carried out.  At the end of the sector
decomposition those integrations that lead to $1/\varepsilon$ poles
and logarithms can be performed by means of standard formulas, which
we derive in the appendices, using recursive subtractions of the type
\refeq{onedimexample1} and \refeq{onedimexample2}.

In this section we describe the sector-decomposition technique to
extract $1/\varepsilon$ poles from massless%
\footnote{This technique is applicable also to massive diagrams,
  however, it permits to isolate only the $1/\varepsilon$ poles and
  not the mass-singular logarithms.}  diagrams \cite{Binoth:2000ps}.
A sector-decomposition algorithm to extract logarithmic singularities
from massive diagrams is presented in \refse{se:massivedecomp}.

\subsection{Primary sector decomposition}
\label{se:primary}

As discussed in \refse{se:subtraction}, our capability to isolate and
simplify the singularities relies on the fact that they originate from
regions of the type \refeq{singularregions}.  Therefore, when one
eliminates the $\delta$-function from the FP integrals
\refeq{fpoutput}--\refeq{fpoutput2} by performing the first
integration, care must be taken that
the structure of the singular regions remains unchanged.%
\footnote{ In particular, one cannot integrate one of the FPs, let's
  say $\alpha_i$, down to $\alpha_i=0$. Otherwise, a singularity that
  might be located in the hyperplane $\{\vec\alpha |\alpha_i=0 \}$
  would be shifted to the hyperplane $\{\vec\alpha|\sum_{j\neq
    i}\alpha_j=1\}$.}  
To this end, one can decompose the $(I-1)$-dimensional hyperplane
that is defined by the
$\delta$-function, into $I$ primary sectors
$\mathcal{\psect}_1,\dots,\mathcal{\psect}_I$, defined as
$\mathcal{\psect}_i=\{\vec{\alpha}\,|\,\alpha_k\le \alpha_i$ for $ 1\le k
\le I $ and $ \sum_{j=1}^I\alpha_j=1 \}$, 
such that the sector $\mathcal{\psect}_i$ does not contain the
hyperplane $\{\vec\alpha |\alpha_i=0\}$.  This yields
\beqar
G&=&\sum_{i=1}^I G_i\qquad
\mbox{with}
\quad
G_i=\int_0^1 \rd \alpha_i \int_0^{\alpha_i} 
\prod_{k=1\atop k\neq i}^I \rd \alpha_k \,
\de\left( 1- \sum_{j=1}^I \alpha_j \right)
\alpha_i^{-I}\vghat\left(\frac{\vec{\alpha}}{\alpha_i}\right)
,
\eeqar
where we used the fact that $\vghat$ is a homogeneous function of
degree $-I$ in the integration variables, \ie $\vghat(\lambda
\vec{\alpha})=\lambda^{-I}\vghat(\vec{\alpha})$ as can be easily seen
from \refeq{feynparamformula}.  Now, each sector $\mathcal{\psect}_i$
can be mapped to the $(I-1)$-dimensional hypercube by means of the
variable transformation
\beqar\label{primaryvtrans}
\alpha_k&=&\left\{
\begin{array}{cl}
\alpha_i \eta_k & \mbox{  if}\quad 1\le k <i \\
\alpha_i \eta_{k-1} & \mbox{  if}\quad i<k\le I
\end{array}
\right.
\quad\mbox{with}\quad
\rd^{I}\vec{\alpha}=\rd \alpha_i\, \alpha_i^{I-1}\rd^{I-1}\vec{\eta}.
\eeqar
Then, eliminating the $\delta$-function through the
$\alpha_i$-integration one obtains
\beqar\label{primarysectint}
G_i &=&
\int_0^1 \frac{\rd \alpha_i}{\alpha_i} 
\int_0^1  \mass{I-1}{\eta}{}\,
\de\left( 1- \alpha_i\left(1+\sum_{j=1}^{I-1} \eta_j \right)\right)
\vghat_i(\vec{\eta})
=
\int_0^1  \mass{I-1}{\eta}{}\,
\vghat_i(\vec{\eta}),
\eeqar
where 
\beq
\vghat_i(\vec{\eta})=\vghat\left(\frac{\vec{\alpha}}{\alpha_i}\right)=\vghat(\eta_1,\dots,\eta_{i-1},1,\eta_{i},\dots,\eta_{I-1}),
\eeq
and using analogous definitions for $\Udet_i$, $\F_i$ and  $\numerator_i$, 
\beqar
\vghat_i(\vec \eta)&=&
\Gamma(\esp+L\varepsilon)
\frac{\numerator_i(\vec{\eta})}
{\left[\F_i(\vec{\eta})\right]^{\esp+L\varepsilon}
\left[\Udet_i(\vec{\eta})\right]^{f-(L+1)\varepsilon}}
.
\eeqar

It is easy to see that this primary sector decomposition does not
modify the structure of the singular regions \refeq{singularregions}.
Indeed, each hyperplane $\{\vec\alpha|\alpha_k=0$ and
$\sum_{j=1}^I\alpha_j=1 \}$, which can give rise to singularities, has
been divided into the sectors $\mathcal{\psect}_i$ with $i\neq k$, and
there it has been mapped to the hyperplanes
$\{\vec\eta|\eta_l=\alpha_k/\alpha_i=0\}$ with $l=k$ for $i>k$ and
$l=k-1$ for $i<k$.  Therefore, all singularities still originate from
regions of the type $\{\vec{\eta}\,|\,\eta_{j_1}=\dots=\eta_{j_n}=0\}$
with $1\le n \le I-1$.

\subsection{Iterated decomposition of UV and mass singularities}
\label{se:iteateddecomp}

Now we perform an iterative decomposition of the primary-sector
integrals \refeq{primarysectint} into sums of subsector integrals,
where all FPs $\eta_i$ that give rise to singularities can be
factorized.  The UV singularities associated with the polynomial
$\Udet_i$ are decomposed as follows:
\begin{enumerate}
  
\item If $\Udet_i(\vec 0)=0$, then we proceed with steps 2--4.
  Otherwise no further decomposition is performed.

\item We choose a set 
$\mathcal{\fpset}=\{\eta_{j_1},\dots,\eta_{j_r}\}$
with a minimal number $r$ of FPs such that  
\beq\label{sectchoiche}
\Udet_i(\vec{\eta})=0
\quad
\mbox{if}
\quad
\eta_{j_1}=\dots = \eta_{j_r}=0.
\eeq
Here, in order to simplify the notation, we assume that $j_k=k$ such
that the polynomial $\Udet_i$ can be written as
\beqar\label{factoform}
\Udet_i(\vec{\eta})=\sum_{k=1}^r \eta_{k} \,
\hat{\Udet}_{ik}(\vec{\eta}).
\eeqar

\item The integration range is decomposed into $r$ subsectors
$\mathcal{S}_1,\dots,\mathcal{S}_r$ with 
$\mathcal{S}_j=\{\vec{\eta}\,|\,\eta_{k}\le \eta_{j}$ for $ 1\le k \le r\}$, 
so that
\beqar
G_i&=&\sum_{j=1}^r G_{ij}\quad
\mbox{with}
\quad
G_{ij}=
\int_0^1  \rd \eta_{j}
\int_0^{\eta_{j}}  \prod_{k=1\atop k\neq j}^r \rd \eta_{k} 
\int_0^1  \prod_{l=r+1}^{I-1} \rd \eta_{l}
\,\vghat_i(\vec{\eta})
.
\eeqar
Each sector $\mathcal{S}_j$ is remapped to the unit cube using the
variable transformation
\beqar\label{secvtrans}
\eta_k&=&\left\{
\begin{array}{cl}
\xi_k & \mbox{if } r+1\le k \le I-1 
\\
\xi_j & \mbox{if } k=j
\\
\xi_j \xi_k & \mbox{otherwise}
\end{array}
\right.
\quad\mbox{with}\quad
\mass{I-1}{\eta}{}=
\mass{I-1}{\xi}{}\, \xi_j^{r-1}.
\eeqar
As a consequence, in the sector $\mathcal{S}_j$ the variable $\xi_j$
can be factorized by rewriting the polynomial $\Udet_i$ as
\beqar
\Udet_i(\vec{\eta})=\xi_j\Udet_{ij}(\vec{\xi})
\quad\mbox{with}\quad
\Udet_{ij}(\vec{\xi})=
\hat{\Udet}_{ij}(\vec{\eta})
+\sum_{k=1\atop k\neq j}^r \xi_k 
\hat{\Udet}_{ik}(\vec{\eta}).
\eeqar
The resulting sector integrals read
\beq \label{Udecsubint}
G_{ij}=\int_0^1\mass{I-1}{\xi}{}\;
\xi_j^{r-1-f+(L+1)\varepsilon}
\vghat_{ij}(\vec{\xi}),
\eeq
where
\beqar
\vghat_{ij}(\vec{\xi})&=&
\Gamma(\esp+L\varepsilon)
\frac{\numerator_{ij}(\vec{\xi})}
{\left[\F_{ij}(\vec{\xi})\right]^{\esp+L\varepsilon}
\left[\Udet_{ij}(\vec{\xi})\right]^{f-(L+1)\varepsilon}}
\eeqar
with%
\footnote{In some cases the FP $\xi_j$ can be factorized also in
  $\F_{ij}(\vec{\xi})$ or $\numerator_{ij}(\vec{\xi})$.}
$\F_{ij}(\vec{\xi})=\F_{i}(\vec{\eta})$ and
$\numerator_{ij}(\vec{\xi})=\numerator_{i}(\vec{\eta})$.

\item For each subsector integral $G_{ij}$, we restart the
  decomposition from step 1.

\end{enumerate} 
The iterative application of steps 1--4 gives rise to a tree-like
structure.  At each iteration the subsectors are divided into new
subsubsectors, which have to be labelled with new indices
\beq
G_{i}\to\sum_{j_1}G_{ij_1},\qquad
G_{ij_1} \to \sum_{j_2}G_{ij_1j_2},\qquad\dots\qquad
G_{ij_1\dots j_{n-1}} \to \sum_{j_{n}}G_{ij_1\dots j_{n-1}j_n}.
\eeq
At each decomposition new FPs are factorized and the 
resulting sector integrals have the general form 
\beq\label{stdsdout1b}
G_{ij_1\dots}=
\int_0^1\mass{I-1}{\xi}{}
\,\product{\vec{\vT}_{ij_1\dots}+\varepsilon\vec{\vtau}_{ij_1\dots}}{\xi}{}
\vghat_{ij_1\dots}(\vec{\xi})
,
\eeq
where 
\beqar\label{stdsdout2b}
\vghat_{ij_1\dots}(\vec{\xi})&=&
\Gamma(\esp+L\varepsilon)
\frac{\numerator_{ij_1\dots}(\vec{\xi})}
{\left[\F_{ij_1\dots}(\vec{\xi})\right]^{\esp+L\varepsilon}
\left[\Udet_{ij_1\dots}(\vec{\xi})\right]^{f-(L+1)\varepsilon}}
,
\eeqar
and the factorized FPs are written as
$\product{\vec{\vT}_{ij_1\dots}+\varepsilon\vec{\vtau}_{ij_1\dots}}{\xi}{}$
using the shorthands
\refeq{productdefinition}--\refeq{exponentdefinition}.  The iteration
of steps 1--4 stops when only sector integrals with
$\Udet_{ij_1\dots}(\vec{0}) \neq 0$ remain.  Then, the same iterative
decomposition has to be repeated for the mass singularities that are
associated with the polynomial $\F_{ij_1\dots}$ until
\beq\label{stdsdout3b}
\Udet_{ij_1\dots}(\vec{0}) \neq 0\quad\mbox{and}\quad
\F_{ij_1\dots}(\vec{0})\neq 0
\eeq
in all subsectors.

The convergence and the efficiency of the sector-decomposition
algorithm depend on the choice of the sets of FPs
$\mathcal{\fpset}=\{\eta_{j_1},\dots,\eta_{j_r}\}$ in step 2, which
determines the subsectors $\mathcal{S}_1,\dots,\mathcal{S}_r$ in step
3.  In principle, the choice of minimal sets $\mathcal{\fpset}$, \ie
sets with a minimal number $r$ of FPs, permits to reduce the number of
new subsectors that are generated at each iteration and guarantees a
certain efficiency.  However, there are in general different minimal
sets $\mathcal{\fpset}$ that fulfil \refeq{sectchoiche}, and it is not
clear how to choose between them in order to minimize the total number
of subsectors at the end of the iterated decomposition.  Moreover, in
some cases it is possible that owing to an unfortunate choice of the
minimal sets the conditions \refeq{stdsdout3b} are never realized, \ie
the convergence of the algorithm, as it is formulated above, is not
guaranteed.  The approach that we have adopted to avoid this problem
consists in a random choice of the minimal sets $\mathcal{\fpset}$.
This strategy is in general not optimal but has the advantage to be
non-deterministic, in the sense that every time that a certain
integral is computed it is decomposed into different sums of subsector
integrals, and this allows for checks on the final result.

\subsection{Counting of singularities}
\label{se:masslesscounting}

In the following we focus on the properties of a generic sector
integral.  To this end we define
\beq\label{stdsdout1}
\vGt(\vec\vT,\espt,\vgt,\Ft)=
\int_0^1\mass{I-1}{\xi}{}
\,\product{\vec{\vT}+\varepsilon\vec{\vtau}}{\xi}{}
\frac{\vgt(\vec{\xi})}
{\left[\Ft(\vec{\xi})\right]^{\espt+L\varepsilon}}
\eeq
with
\beqar\label{stdsdout2}
\vgt(\vec{\xi})&=&
\Gamma(\esp+L\varepsilon)
\frac{\numeratort(\vec{\xi})}
{\left[\Udett(\vec{\xi})\right]^{f-(L+1)\varepsilon}}
\eeqar
and
\beq\label{stdsdout3}
\Udett(\vec{0}) \neq 0,\qquad
\Ft(\vec{0})\neq 0 
.
\eeq
The sector integrals \refeq{stdsdout1b} can be written as
$G_{ij_1\dots}=\vGt(\vec\vT_{ij_1\dots},\esp,\vg_{ij_1\dots},\F_{ij_1\dots})$
with these conventions.  The dependence of $\vGt$ on $\vec\vT$,
$\espt$, $\vgt$ and $\Ft$ has been kept in explicit form for later
convenience, whereas the dependence on all other quantities is
implicitly understood. Note also that $\espt$ can be different from
$\esp$ in \refeq{stdsdout1}--\refeq{stdsdout2}.

The FPs that give rise to singularities in the sector integrals
\refeq{stdsdout1}--\refeq{stdsdout3} can be identified by simple power
counting.  In principle all FPs in $\mathcal{S}=\{\xi_k | \vT_k\le
-1\}$ yield singularities unless these are not compensated by the
behaviour of $\numeratort(\vec \xi)$ at $\vec \xi \to \vec 0$.
Therefore, at the end of the sector decomposition, we decompose
$\numeratort(\vec \xi)$ into monomials with respect to the FPs in
$\mathcal{S}$ and we treat the contribution from each monomial as a
different subsector, such that the FPs $\mathcal{S}$ can be factorized
and the resulting numerator $\numeratort$ does not depend on them
anymore.
 
Now, those FPs that do and those that do not give rise to
singularities can be immediately identified.  If we rename them as
$\vec{y}=(y_1,\dots,y_m)$ and $\vec{x}=(x_1,\dots,x_l)$, respectively,
with $l+m=I-1$, then the contribution from each subsector assumes the
general form
\beqar\label{genmasslesssdout}
\vGt &\equiv& \vGt(\vec\vT,\espt,\vgt,\Ft)=
\int_0^1 \mass{\vl}{\vx}{}
\,\product{\vec{\vA}+\vec{\valpha}\varepsilon}{\vx}{}
\int_0^1 \mass{\vm}{\vy}{}
\,\product{\vec{\vbeta}\varepsilon-\vec{\vb}-1}{\vy}{}
\frac{\vgt(\vec{\vx};\vec{\vy})}
{\left[\Ft(\vec{\vx};\vec{\vy})\right]^{\espt+L\varepsilon}}
\eeqar
with 
\beqar\label{genmasslesssdout2}
\vgt(\vec{\vx};\vec{\vy})
&=&
\Gamma(\esp+L\varepsilon)
\frac{\numeratort(\vec{\vx})}
{\left[\Udett(\vec{\vx};\vec{\vy})\right]^{f-(L+1)\varepsilon}}
\eeqar
and exponents
\beqar\label{masslessexp}
\vec \vA &=&(\vA_1,\dots,\vA_l),\qquad 
\vec \valpha =(\valpha_1,\dots,\valpha_l),\nl
\vec \vb &=&(\vb_1,\dots,\vb_m),\qquad 
\vec \vbeta =(\vbeta_1,\dots,\vbeta_m).
\eeqar
By construction, all exponents \refeq{masslessexp} are integer numbers
with $\vA_i> -1$ and $\vb_j\ge 0$.  The exponents $\vec{\vA}$,
$\vec{\vb}$ and $\vec{\valpha}$, $\vec{\vbeta}$ are related to
$\vec{\vT}$ and $\vec{\vtau}$, respectively, in an obvious way.  As
mentioned before, we have decomposed $\numeratort(\vec\vx,\vec\vy)$
into monomials with respect to the FPs $\vec\vy$ and we treat the
contribution from each monomial as a different subsector.  The
integrations over the $\vm$ FPs $\vec\vy$ give rise to singularities
up to the order $1/\varepsilon^m$.

\subsection{Integration}
The integrations of the FPs $\vy_1,\dots,\vy_\vm$ for a generic
integral of the type \refeq{genmasslesssdout} are performed in
\refapp{se:onescaleint}.  This is done through recursive subtractions
of the type \refeq{onedimexample1}, which permit to extract the
leading and next-to-leading singularities, \ie the poles of order
$1/\varepsilon^m$ and $1/\varepsilon^{m-1}$.  The result for the
integral \refeq{genmasslesssdout} is obtained by applying
\refeq{onescaleformulalogsing} to the integrand
\beq
{\fun}(\vec{\vy})
=
\int_0^1\mass{\vl}{\vx}{}\,
{\fun}(\vec{\vx};\vec{\vy})
\quad\mbox{with}\quad
\fun(\vec{\vx};\vec{\vy})
=
\product{\vec{\vA}+\vec{\valpha}\varepsilon}{\vx}{}
\frac{\vgt(\vec{\vx};\vec{\vy})}
{\left[\Ft(\vec{\vx};\vec{\vy})\right]^{\espt+L\varepsilon}}
.
\eeq
Expanding $\fun\equiv\fun(\vec{\vx};\vec{\vy})$
in $\varepsilon$ as
\begin{equation}\label{epsilonexp1}
\fun\equiv
\sum_{\vj= 0}^\infty \frac{\varepsilon^\vj}{\vj!}\fun^{(\vj)},
\end{equation}
and similarly%
\footnote{ If $\esp\le 0$, then the leading term of
  $\Gamma(\esp+\varepsilon)\equiv \sum_{\vj=0}^{\infty}
  \Gamma^{(\vj)}(\esp)\varepsilon^\vj/\vj!$ contains a $1/\varepsilon$
  pole:
\begin{displaymath}
\Gamma^{(0)}(\esp)=
\frac{(-1)^\esp}{(-\esp)!}\frac{1}{\varepsilon}\quad
\mbox{for}\quad \esp\le 0.
\end{displaymath}
}
$\Gamma(\esp+\varepsilon)$ and $\numeratort$
we obtain
\beqar\label{masslessintegrationresult}
\vGt
&\NLL&
\left[ 
\product{}{\vbeta}{}
\right]^{-1}
\left(\frac{1}{\varepsilon}\right)^{\vm}
\int_0^1\mass{\vl}{\vx}{}
\Biggl\{
\deriv^{\vec{\vb}}_{\vec{\vy}}\,{\fun^{(0)}(\vec{\vx};\vec{0})}
\nl&&{}
+\varepsilon
\left[
\deriv^{\vec{\vb}}_{\vec{\vy}}\,{\fun^{(1)}(\vec{\vx};\vec{0})}
+\sum_{\vi=1}^\vm \vbeta_\vi \Delta_{\vec\vy}^{\vec\vb,\vi}
{\fun^{(0)}(\vec{\vx};\vec{0})}
\right]
\Biggr\}
\eeqar
to NLL accuracy with
\beqar\label{mleadsubleadexp}
\fun^{(0)}
&=&
\Gamma^{(0)}(\esp) 
 \product{\vec{\vA}}{\vx}{}
\frac
{\numeratort^{(0)}}
{\Ft^{\espt}\,
\Udett^{f}} 
,\nl
\frac{\fun^{(1)}
}
{\fun^{(0)}
}
&=&
L\frac{\Gamma^{(1)}(\esp)}{\Gamma^{(0)}(\esp)}
+\log\left[
\product{\vec{\valpha}}{\vx}{}
\right]
+\frac{\numeratort^{(1)}
}{
\numeratort^{(0)}
}
+(L+1)
\log
\Udett
-L\log
\Ft
.
\eeqar
The derivative operators $\deriv^{\vec{\vb}}_{\vec{\vy}}$ and the
subtraction operators $\Delta^{\vec{\vb},i}_{\vec{\vy}}$ are defined
in \refeq{multiderivdef} and \refeq{multimasssubdef}, respectively.
These latter are associated to those next-to-leading contributions
that remain after subtraction of the singularity from the $\vy_\vi$
integration.  For instance, in the special case $\vec{\vb}=\vec{0}$
where all $\vec{\vy}$ integrations are logarithmically singular, we
simply have
\beqar
\Delta_{\vec\vy}^{\vec 0, \vi}
\fun^{(0)}(\vec{\vx};\vec{0})
&=&\int_0^1\frac{\rd \vy_{\vi}}{\vy_{\vi}}
\left[
\fun^{(0)}(\vec{\vx};0,\dots, \vy_{\vi},\dots,0)
-\fun^{(0)}(\vec{\vx};\vec{0})
\right].
\eeqar
By construction, the remaining $\vy_\vi$ integration in the
$\Delta^{\vec{\vb},i}_{\vec{\vy}}$ term is finite and can be performed
either numerically or analytically.  Also the $\vl$-dimensional
integral over the FPs $\vec \vx$ in \refeq{masslessintegrationresult}
is convergent.  The coefficients of the leading and next-to leading
poles have thus been expressed in terms of convergent integrals with
dimension $\vl$ and $\vl+1$.  For negative kinematical invariants
these integrals are well-defined.  For positive kinematical invariants
the analytic continuation is provided by the infinitesimal imaginary
part $\epsilon$ contained in $\Ft$.

\section{Sector decomposition of massive diagrams}
\label{se:massivedecomp}

Let us now consider the sector decomposition of massive FP integrals
\refeq{fpoutput}--\refeq{fpoutput2} assuming a hierarchy of energy and
mass scales as in \refeq{scales}.  In order to extract the
singularities that arise as combinations of $1/\varepsilon$ poles and
logarithms of $\ratio=M^2/s$, which appear in the asymptotic limit
$\ratio\ll 1 $, we proceed as follows.

First, we perform the primary and iterated sector decompositions as
described in \refses{se:primary}--\ref{se:iteateddecomp}.  As a
result, the sector integrals assume the form
\refeq{stdsdout1}--\refeq{stdsdout3}, where owing to \refeq{stdsdout3}
all FPs which lead to $1/\varepsilon$ poles are factorized.  Then, in
each sector, we decompose the polynomial $\Ft$ as
\beqar\label{eq:Ft_decomp}
\Ft(\vec{\xi})=
\vfs(\vec{\xi})+
\ratio \vfm(\vec{\xi})-\ri\epsilon,
\eeqar
where $\vfs(\vec\xi)$ and $\vfm(\vec\xi)$ are the contributions
associated with the high-energy and mass parameters, and depend
linearly on the corresponding ratios $s_j/s$ and $M_k^2/M^2$,
respectively. If $\vfs(\vec 0)=0$, there can be mass singularities
that are regulated by the mass term $\ratio \vfm(\vec\xi)$ giving rise
to logarithms of $\ratio$ in the asymptotic limit.  In this case, we
perform an additional iterated decomposition analogous to that of
\refse{se:iteateddecomp} until all FPs that lead to $\vfs(\vec 0)=0$
are factorized.  At the end the original integral consists of a sum of
sector integrals of the type
\beq\label{cstdsdout1}
\vGt(\vec\vT,\espt,\vgt,\Ft)=\int_0^1\mass{I-1}{\xi}{}
\,\product{\vec{\vT}+\vec{\vtau}\varepsilon}{\xi}{}
\frac{\vgt(\vec{\xi})}{\left[\Ft(\vec{\xi})\right]^{\espt+L\varepsilon}}
\eeq
where
\beqar\label{cstdsdout2}
\vgt(\vec{\xi})&=&
\Gamma(\esp+L\varepsilon)
\frac{\numeratort(\vec{\xi})}
{\left[\Udett(\vec{\xi})\right]^{f-(L+1)\varepsilon}}
,
\eeqar
and 
\beqar\label{decden}
\Ft(\vec{\xi})=
\,\product{\vec{\vt}}{\xi}{}
\vfst(\vec{\xi})+
\ratio  \vfmt(\vec{\xi})-\ri\epsilon.
\eeqar
After the last sector decomposition we have 
\beq\label{cstdsdout3}
\Udett(\vec{0}) \neq 0,\qquad 
\Ft(\vec{0})\neq 0, 
\qquad\mbox{and}\quad \vfst(\vec{0})\neq 0 
\eeq
in all subsectors. Without loss of generality we can assume that%
\footnote{This requires a permutation of the FPs that is, in general,
  different for different subsectors.}
\beqar\label{factparexp}
\vec \vt=(\vt_1,\dots\vt_p,0,\dots,0)\qquad
\mbox{with}\quad \vt_i>0 \quad\mbox{for}\quad i=1,\dots,\vp,
\eeqar
\ie that the FPs that have been factorized in 
\refeq{decden} are the first $p$ FPs, 
where $0\le \vp \le I-1$.

Simple power counting indicates that the set of FPs which can give
rise to $1/\varepsilon$ poles or mass-singular logarithms is given by
$\mathcal{S}=\{\xi_k| \vT_k\le -1$ or $ \vT_k-\espt \vt_k \le -1\}$.
As discussed in \refse{se:masslesscounting}, the numerator
$\numeratort(\vec \xi)$ has to be decomposed into monomials with
respect to the FPs in $\mathcal{S}$, and each monomial has to be
treated as a different subsector where the FPs in $\mathcal{S}$ can be
factorized.

\subsection{Transformation of non-logarithmic singularities}
\label{se:remapping}
Let us discuss the behaviour of the integrations over the FPs
$\xi_1,\dots,\xi_\vp$ in the asymptotic limit $\ratio\to 0$.  If
$\vT_k>-1$ for $1\le \vk\le \vp$, the $\xi_k$-integrations are
convergent in $D=4$ dimensions.  For $\ratio\to 0$ they behave as
\beqar
\int \rd \xi_k \frac{\xi_k^{\vT_k}}{\left[\xi_k^{\vt_k}+\ratio \right]^\espt}
\sim
\int \rd \xi_k^{\vt_k} \frac{(\xi_k^{\vt_k})^{\espt-d_k-1}}{\left[\xi_k^{\vt_k}+\ratio \right]^\espt}
\sim \left\{
\begin{array}{cl}
1&\mbox{  if    } d_k<0 \\
\log(\ratio)&\mbox{  if    } d_k=0 \\
\ratio^{-d_k}&\mbox{  if    } d_k>0
\end{array}
\right.
,
\eeqar
where we have introduced the degree of singularity 
\beq\label{singdegdef}
d_k:=\espt-\frac{\vT_k+1}{\vt_k}
\eeq
for $k=1,\dots,\vp$. Note that if $d_k>0$, the $\xi_k$-integral yields
a non-logarithmic mass singularity of order $\ratio^{-d_k}$ for
$\vT_k>-1$.  Similarly, for $\vT_k\le -1$ one obtains singular
contributions of order $\ratio^{-d_k} \varepsilon^{-1}$ in
$D=4-2\varepsilon$ dimensions.  Independently of the values of
$\vT_k$, these non-logarithmically divergent integrals can be easily
expressed as a linear combination of logarithmically-divergent
integrals with $d_k\le 0$ for $k=1,\dots,\vp$.  To this end, we define
the maximal degree of singularity
\beq \label{maxsingdeg}  
d:=\max_{1\le k \le p}(d_k),
\eeq
and if $d>0$ we multiply the integrand in \refeq{cstdsdout1} by
\beqar
1&=&\left[
\frac{\Ft
-\vfst
\product{\vec \vt}{\xi}{}
}
{\ratio \vfmt-\ieps}
\right]^d
=
\left({\ratio\vfmt-\ieps}\right)^{-d}
\sum_{\vr=0}^d\left({d\atop \vr}\right)
\Ft^{d-\vr}
\left[-
\vfst
\product{\vec \vt}{\xi}{}
\right]^\vr.
\eeqar
This yields
\beqar\label{lincomb}
\vGt(\vec\vT,\espt,\vgt,\Ft)&=&
\ratio^{-d}
\sum_{\vr=0}^d\left({d\atop \vr}\right)(-1)^\vr 
\vGt(\vec\vT_\vr,\espt_\vr,\vgt_\vr,\Ft),
\eeqar
where 
\beqar\label{nonlogtransf}
\vec\vT_\vr=\vec\vT+\vr \vec\vt
,\qquad
\espt_\vr=\espt-d+\vr
,\qquad
\vgt_\vr=\vgt (\vfst)^\vr (\vfmt-\ieps)^{-d}
.
\eeqar
As a result of the above transformation, the term $\ratio^{-d}$
involving inverse powers of masses is factorized and the degree of
divergence $d_k$ is reduced by $d$ in the resulting integrals
$\vGt(\vec\vT_\vr,\espt_\vr,\vgt_\vr,\Ft)$.  This can be easily seen
from
\beqar
\left(\vec d_\vr\right)_k&:=&
\espt_r-\frac{\left(\vec\vT_\vr\right)_k+1}{\vt_k}=
d_k-d.
\eeqar
Thus, the integrations with maximal degree of singularity $d_k=d$
become logarithmically divergent, whereas all other integrations with
$d_k< d$ become non-divergent.

Since the integrals resulting from
\refeq{lincomb}--\refeq{nonlogtransf} have the same structure as the
original ones, we need to compute only integrals of the type
\refeq{cstdsdout1}--\refeq{factparexp} with
\beqar\label{dcase}
d_k\le d \le 0\quad\mbox{for}\quad k=1,\ldots,p
\eeqar
in the following. Integrals with $\vT_\vk\le -1$ for some $1\le k \le
\vp$, where the corresponding $\xi_k$-integrations give rise to
additional $1/\varepsilon$ poles, do not need to be computed
explicitly\footnote{In practical calculations, we never encountered
  contributions of this type at the NLL level.  However, in general we
  cannot exclude their presence, especially in view of an extension of
  the algorithm to the NNLL level or beyond.  } since, as we show in
\refapp{se:ibp}, this kind of integrals can be eliminated by means of
integration-by-parts identities.  Thus, we can restrict ourselves to
integrals with
\beqar\label{Tcase}
\vT_k>-1\quad \mbox{for}\quad k=1,\dots,p.
\eeqar
We note that, as a consequence of \refeq{dcase}--\refeq{Tcase},
\beq\label{ecase}
d=0 \quad \Rightarrow \quad \espt >0,
\eeq
since $d=0$ implies that there is a $k$ with $1\le k\le \vp$ and
$d_k=0$. Thus $\espt=(\vT_k+1)/\vt_k>0$ owing to  \refeq{singdegdef} and  \refeq{Tcase}.
\subsection{Classification and counting of singularities}
\label{se:classif}
In the following, we consider sector integrals of the type
\refeq{cstdsdout1}--\refeq{factparexp} with
\refeq{dcase}--\refeq{Tcase}.  In order to extract the various
$1/\varepsilon$ poles and $\log(\ratio)$ contributions, we first
classify the FP integrations in \refeq{cstdsdout1}--\refeq{factparexp}
according to their singular behaviour.
\begin{itemize}
\item The FPs 
  $\{\xi_k| 1\le k \le \vp$ and $d_k=0\}$, for which we can assume
  $\vt_k>0$ and $\vT_k>-1$, are renamed as $\{z_1,\dots,z_n\}$.  As we
  see in \refse{se:massiveintegration}, the corresponding $n$
  integrations give rise to singular contributions up to the order
  $\log^{\vn}(\ratio)$.
  
\item The FPs $\{\xi_j| \vp+1\le j \le I-1$ and $\vT_j\le -1 \}$, for
  which we can assume $\vt_j=0$, are renamed as $\{y_1,\dots,y_m\}$.
  The corresponding $m$ integrations give rise to poles up to the
  order $1/\varepsilon^m$.
 
\item The remaining FPs, $\{\xi_i| \vT_i>-1$ and if $1\le i \le \vp,
  d_i<0\}$, are called $\{x_1,\dots,x_l\}$.  The $l$ integrations over
  these FPs are free of singularities.
\end{itemize}
In general we have $0\le l,m,n \le I-1$ and $l+m+n=I-1$.  As a result
of the above classification, the contribution from each subsector
assumes the general form
\beqar\label{gendecmassiveint1}
\lefteqn{
\vGt\equiv\vGt(\vec\vT,\espt,\vgt,\Ft)=
}\quad\nl&=&
\int_0^1 \mass{\vl}{\vx}{}
\,\product{\vec{\vA}+\vec{\valpha}\varepsilon}{\vx}{}
\int_0^1 \mass{\vm}{\vy}{}
\,\product{\vec{\vbeta}\varepsilon-\vec{\vb}-1}{\vy}{}
\int_0^1 \mass{\vn}{\vz}{}
\,\product{\vec\vc \espt-1 +\vec{\vgamma}\varepsilon}{\vz}{}
\frac{\vgt(\vec{\vx};\vec{\vy};\vec{\vz})}
{\left[
\Ft(\vec{\vx};\vec{\vy};\vec{\vz})
\right]^{\espt+L\varepsilon}}
\nln
\eeqar
with
\beqar\label{gendecmassiveint2a}
\vgt(\vec{\vx};\vec{\vy};\vec{\vz})
&=&
\Gamma(\esp+L\varepsilon)
\frac{\numeratort(\vec{\vx})}
{\left[\Udett(\vec{\vx};\vec{\vy};\vec{\vz})\right]^{f-(L+1)\varepsilon}}
\eeqar
and
\beqar\label{gendecmassiveint2}
\Ft(\vec{\vx};\vec{\vy};\vec{\vz})
&=&
\product{\vec\va}{\vx}{}
\product{\vec\vc}{\vz}{}
\vfst(\vec{\vx};\vec{\vy};\vec{\vz})
+\ratio \vfmt(\vec{\vx};\vec{\vy};\vec{\vz})-\ieps
,
\eeqar
where $\Udett(\vec{0};\vec{0};\vec{0})\ne0$,
$\Ft(\vec{0};\vec{0};\vec{0})\neq 0$, and
$\vfst(\vec{0};\vec{0};\vec{0})\neq 0$.  The exponents that are
associated to the FPs $\vec \vx$ and $\vec \vy$,
\beqar
\vec \vA &=&(\vA_1,\dots,\vA_l),\qquad 
\vec \va =(\va_1,\dots,\va_l),\qquad 
\vec \valpha =(\valpha_1,\dots,\valpha_l),
\nl
\vec \vb &=&(\vb_1,\dots,\vb_m),\qquad 
\vec \vbeta =(\vbeta_1,\dots,\vbeta_m),
\eeqar
are integer numbers and satisfy 
\beq\label{massexponents}
\vA_i>-1,\qquad
\vA_i>\va_i \espt -1,\qquad
\va_i\ge 0,\qquad
\vb_j\ge 0,
\eeq
by construction.  The number $\vn$ of FPs $\vec \vz$ depends on the
maximal degree of singularity $d$. If $d<0$ then $\vn=0$, otherwise
$\vn>0$ and for the exponents of the FPs $\vec \vz$,
\beqar
\vec \vc\, \espt &=&(\vc_1 \espt,\dots,\vc_n \espt),\qquad \vec
\vgamma 
=(\vgamma_1,\dots,\vgamma_n),
\eeqar
which are integer numbers, we have 
\beq\label{massexponents2}
\vc_k>0,\qquad
\vc_k \espt> 0.
\eeq
Again, the exponents $\vec{\vA}$, $\vec{\va}$, $\vec{\vb}$,
$\vec{\vc}$ and $\vec{\valpha}$, $\vec{\vbeta}$, $\vec{\vgamma}$ are
related to $\vec{\vT}$, $\vec{\vt}$ and $\vec{\vtau}$ in an obvious
way, $\numerator(\vec{\vx},\vec{\vy},\vec{\vz})$ has been decomposed
into monomials with respect to the FPs $\vec\vy$ and $\vec\vz$ and the
contribution from each monomial has been treated as a different
subsector.  Note that $n>0$, or equivalently $d=0$, implies $\espt>0$,
as already observed in \refeq{ecase}.

\subsection{Integration}
\label{se:massiveintegration}
Before we proceed,
we rewrite 
\refeq{gendecmassiveint1}--\refeq{gendecmassiveint2}
as%
\beqar\label{gendecmassiveint1b}
\vGt&=&
\int_0^1 \mass{\vl}{\vx}{}
\int_0^1
\mass{\vm}{\vy}{}
\,\product{\vec{\vbeta}\varepsilon-\vec{\vb}-1}{\vy}{}
\int_0^1
\mass{\vn}{\vz}{}
\,\vfn(\vec{\vx};\vec{\vy};\vec{\vz})\,
\frac{
\product{\vec{\vc}\espt-1+\vec{\vgamma}\varepsilon}{\vz}{}
}{
\left[
\product{\vec{\vc}}{\vz}{}
+\ratio \vfr(\vec{\vx};\vec{\vy};\vec{\vz})
\right]^{\espt+L\varepsilon}}
\eeqar
with
\begin{equation}\label{h0def}
\vfn(\vec{\vx};\vec{\vy};\vec{\vz})
=
\product{\vec{\vA}-\espt\vec{\va}+(\vec{\valpha}-L\vec{\va})\varepsilon}{\vx}{}
\,\frac{\vgt(\vec{\vx};\vec{\vy};\vec{\vz})}
{\left[\vfst(\vec{\vx};\vec{\vy};\vec{\vz})-\ieps\right]^{\espt+L\varepsilon}}
\end{equation}
and
\beq\label{h1def}
\vfr(\vec{\vx};\vec{\vy};\vec{\vz})=
\frac{\vfmt(\vec{\vx};\vec{\vy};\vec{\vz})-\ieps}
{\left[\vfst(\vec{\vx};\vec{\vy};\vec{\vz})-\ieps\right] 
\product{\vec{\va}}{\vx}{}}.
\eeq
The singular integrations over the FPs $\vec \vy$ and $\vec\vz$ for a
generic integral of the type \refeq{gendecmassiveint1b} are performed
in \refapp{se:twoscaleint} to NLL accuracy.  The resulting
$1/\varepsilon$ poles and $\log(\ratio)$ contributions can be read off
from \refeq{NLLscaleresultd}.  Expanding
$\vfn\equiv\vfn(\vec{\vx};\vec{\vy};\vec{\vz})$ in $\varepsilon$ as
\begin{equation}
\vfn
\equiv \sum_{\vj= 0}^\infty \frac{\varepsilon^\vj}{\vj!}\vfn^{(\vj)}
\end{equation}
and similarly $\Gamma(\esp+\varepsilon)$ and $\numeratort$,
we obtain
\beqar\label{onescaleintegrationresult}
\vGt&\NLL&\left[ 
\product{}{\vbeta}{}
\product{}{\vc}{}
\right]^{-1}
(-1)^{\vn}
\sum_{\vp\ge0}
\left(\frac{1}{\varepsilon}\right)^{\vm-\vp}
\int_0^1\mass{\vl}{\vx}{}
\Biggl\{
\combin_\vp(\vec{\vr})
\deriv^{\vec{\vb}}_{\vec{\vy}}\,
{\vfn^{(0)}(\vec{\vx};\vec{0};\vec{0})}
\logarsymbol{\vn+\vp}\left(\ratio\right)
\nl&&{}
+
\Biggl[
\theta(\vn-1)
\combin_\vp(\vec{\vr})
\deriv^{\vec{\vb}}_{\vec{\vy}}
\left[{\vfn^{(0)}(\vec{\vx};\vec{0};\vec{0})}
\left[\log\left(
\vfr(\vec{\vx};\vec{0};\vec{0})
\right)+C_\espt\right]\right]
\nl&&{}
-\sum_{\vj=1}^\vn \vc_\vj\combin_{\vp}(\vec{\vr}_{[\vj]}) 
\Delta_{\vec\vz}^{\vec 0,\vj} \deriv^{\vec{\vb}}_{\vec{\vy}}\,
{\vfn^{(0)}(\vec{\vx};\vec{0};\vec{0})}
\nl&&{}
+\combin_{\vp-1}(\vec{\vr})
\left[
\deriv^{\vec{\vb}}_{\vec{\vy}}\,
{\vfn^{(1)}(\vec{\vx};\vec{0};\vec{0})}
+\sum_{\vi=1}^\vm \vbeta_\vi \Delta_{\vec\vy}^{\vec\vb,\vi}
{\vfn^{(0)}(\vec{\vx};\vec{0};\vec{0})}
\right]
\Biggr]
\logarsymbol{\vn+\vp-1}\left(\ratio\right)
\Biggr\},
\eeqar
where $\logarsymbol{\vn}$ is defined in \refeq{logsymbdef} and
\beqar\label{mleadsubleadexp2}
\vfn^{(0)}&=&
\Gamma^{(0)}(\esp) 
 \product{\vec{\vA}-\espt\vec{\va}}{\vx}{}
\frac{\numeratort^{(0)}}
{\left(\vfst-\ieps\right)^{\espt}
\Udett^{f}}
,\nl
\frac{\vfn^{(1)}}{\vfn^{(0)}}
&=&
L\frac{\Gamma^{(1)}(\esp)}{\Gamma^{(0)}(\esp)}
+\log\left(
\product{\vec{\valpha}-L\vec{\va}}{\vx}{}
\right)
+\frac{\numeratort^{(1)}}{\numeratort^{(0)}}
+(L+1)\log\Udett
-L\log \left(\vfst-\ieps\right)
.
\eeqar
The derivative operator $\deriv^{\vec{\vb}}_{\vec{\vy}}$ and the
subtraction operators $ \Delta_{\vec\vy}^{\vec\vb,\vi}$ and
$\Delta_{\vec\vz}^{\vec 0,\vj}$ are defined in \refeq{multiderivdef}
and \refeq{multimasssubdef} and \refeq{multimasssubdef2},
respectively.  The constant $C_\espt$ is the sum defined in
\refeq{Cdefinition}.  The vector $\vec{\vr}=(\vr_1,\dots,\vr_\vn)$ has
components
\beq
\vr_\vj=\frac{\vgamma_\vj}{\vc_\vj}-L,
\eeq
and $\vec{\vr}_{[\vj]}$ and $\combin_{\vp}(\vec{\vr})$ 
are  defined in \refeq{compsuppression} and \refeq{combinfun}, respectively.

As observed at the end of \refapp{se:twoscaleint}, the result
\refeq{masslessintegrationresult} for massless integrals, can be
obtained from the general result \refeq{onescaleintegrationresult} for
massive integrals in the special case $\vn=0$.  In the massive case,
for the integrals
\refeq{gendecmassiveint1}--\refeq{gendecmassiveint2}, where the
integrations over the FPs $\vec\vz$ are logarithmically singular, the
$\vm+\vn$ integrations over the FPs $\vec \vy$ and $\vec\vz$ give rise
to singularities up to the order
$\varepsilon^{-\vm+p}\log^{\vn+p}(\ratio)$ with $p\ge 0$, as one can
see in \refeq{onescaleintegrationresult}.  Instead, for the asymptotic
behaviour of non-logarithmically singular integrals we observe the
following.  For an integral \refeq{cstdsdout1}--\refeq{factparexp}
with $d>0$ and \refeq{Tcase}, which is expressed in terms of
logarithmically singular integrals of the type
\refeq{gendecmassiveint1}--\refeq{gendecmassiveint2} using
\refeq{lincomb}--\refeq{nonlogtransf}, each term in the sum
\refeq{lincomb} yields a contribution of order
$\varepsilon^{-\vm+p}\ratio^{-d}\log^{\vn+p}(\ratio)$ with $p\ge0$.
However, from \refeq{nonlogtransf}, \refeq{h0def} and
\refeq{onescaleintegrationresult} one can see that these leading
contributions are independent of $\vr$ and thus cancel in the sum
\refeq{lincomb}, \ie to LL accuracy
\beqar\label{lincombLL}
\lefteqn{
\ratio^{-d}
\sum_{\vr=0}^d\left({d\atop \vr}\right)(-1)^\vr 
\vGt(\vec\vT_\vr,\espt_\vr,\vgt_\vr,\Ft)
}\quad\nl
&\LL& \ratio^{-d}
\vGt(\vec\vT,\espt-d,\vfmt^{-d}\vgt,\Ft)
\sum_{\vr=0}^d\left({d\atop \vr}\right)(-1)^\vr
=0 
.
\eeqar
Consequently, only the subleading contributions proportional to
$C_{\espt_\vr}$ survive, \ie contributions of order
$\varepsilon^{-\vm+p}\ratio^{-d}\log^{\vn+p-1}(\ratio)$ with $p\ge 0$.

The result \refeq{onescaleintegrationresult} is expressed as
$\vl$-dimensional integral over the FPs $\vec \vx$.  Analogously to
the massless case, the next-to-leading contributions involve one
additional integration over the FPs ${\vy}_{\vi}$ or ${\vz}_{\vj}$,
which appears in the subtraction operator $
\Delta_{\vec\vy}^{\vec\vb,\vi}$ or $\Delta_{\vec\vz}^{\vec 0,\vj}$,
respectively.  Thus, the coefficients of the LLs and NLLs have been
expressed in terms of convergent integrals with dimension $\vl$ and
$\vl+1$.  For negative kinematical invariants these integrals are
well-defined.  For positive kinematical invariants the analytic
continuation is provided by the infinitesimal imaginary part $\ieps$
contained in $\vfn$ and $\vfr$.

\section{Discussion}
\label{se:discussion}
The sector-decomposition algorithm that we have presented in
\refses{se:masslessdecomp} and \ref{se:massivedecomp} permits to
reduce arbitrary massive (or massless) FP integrals to a sum of sector
integrals of the type
\refeq{gendecmassiveint1}--\refeq{gendecmassiveint2}, where all FPs
that can lead to UV or mass singularities are factorized.  Each sector
integral gives rise to a tower
\beqar\label{tower2}
\sum_{j=0}^{J}\sum_{k=-j}^\infty
a_{j,k}\, \varepsilon^{k}  \log^{j+k}(\ratio),
\eeqar
involving $1/\varepsilon$ poles and logarithms.  As discussed in
\refse{se:classif}, the FPs $\vy_1,\dots,\vy_\vm$ and
$\vz_1,\dots,\vz_\vn$, which give rise to these singular contributions
are easily identified by power counting.  The maximal total power $J$
of the resulting poles and logarithms is given by the number of these
FPs $\vec \vy$ and $\vec \vz$, \ie $J=\vm+\vn$.  This simple counting
permits to select the sector integrals that give rise to the LLs and
NLLs by requiring that $J=2L$ and $J\ge 2L-1$, respectively, at
$L$-loop level.  Here we have assumed that the overall factor
$\Gamma(\esp+L\varepsilon)$ in \refeq{fpoutput1} does not yield
$1/\varepsilon$ poles since the LLs and NLLs originate from diagrams
involving mass singularities, \ie diagrams with an exponent $\esp > 0$
for the mass-dependent denominator $\F$ in \refeq{fpoutput1}.

The singular contributions have been extracted from the integrals over
the FPs $\vec\vy$ and $\vec\vz$ in
\refeq{gendecmassiveint1}--\refeq{gendecmassiveint2} using the
subtraction technique discussed in \refse{se:subtraction} and
\refapp{se:twoscaleint} [see \refeq{intexpa}--\refeq{intexp}].  Each
of these $J$ integrations has been split into a singular part that has
been solved analytically (giving rise to a pole or a logarithm) and a
non-singular part that has not been integrated in
\refeq{onescaleintegrationresult}.  The contributions with total power
$j$ in the resulting tower \refeq{tower2} originate from the
combinations of $j'$ singular parts and $J-j'$ non-singular parts with
$j\le j'\le J$.  Their coefficients are thus expressed as integrals
over $(J-j')$ remaining FPs $\vy_\vi$ or $\vz_\vk$.  Finally, together
with these FPs, also the FPs $\vx_1,\dots,\vx_\vl$, which do not give
rise to singularities, remain to be integrated.  In the general result
\refeq{onescaleintegrationresult}, the coefficients of the leading
($j=J$) and next-to-leading ($j=J-1$) contributions for the generic
sector integral \refeq{gendecmassiveint1}--\refeq{gendecmassiveint2}
are expressed as  $(I-1-J)$- and $(I-J)$-dimensional integrals,%
\footnote{The additional integration for the coefficients of the NLLs
  is associated with the subtraction operators $
  \Delta_{\vec\vy}^{\vec\vb,\vi}$ and $\Delta_{\vec\vz}^{\vec 0,\vj}$
  defined \refeq{multimasssubdef}.  } where $I=\vl+\vm+\vn+1$ is the
number of internal lines of the Feynman diagram.  These remaining
integrations are free of UV and mass singularities by construction.

For the L-loop corrections to processes with $E$ external legs, which
involve a maximal number $I=E+3L-3$ of internal lines, the
coefficients of the LLs ($j=2L$) and NLLs ($j=2L-1$) consist of
$(E+L-4)$- and $(E+L-3)$-dimensional integrals, respectively.  For the
most complicated Feynman diagrams that we have computed up to now
\cite{Denner:2003wi,Pozzorini:2004rm}, which correspond to 3- and
4-point functions at two loops, the NLLs require 2- and 3-dimensional
integrals, respectively, and such integrals turn out to be
sufficiently simple to be solved analytically by standard computer
algebra programs such as \Mathematica \cite{Mathematica}.

To our knowledge, the algorithm we have presented is the only existing
tool that permits to extract NLLs from arbitrary Feynman diagrams in a
completely automated way.  At present, the accuracy is limited only by
the fact that the divergent integrals have been solved in NLL
approximation.  In order to extend the algorithm beyond the NLL level
it is sufficient to solve these divergent integrations, which
correspond to a well-defined class of standard integrals resulting
from the sector decomposition, with higher accuracy.

The algorithm has been checked against all one-loop tensor integrals
given in \citere{Roth:1996pd} and the two-loop master integrals of
\citere{Aglietti:2003yc}.  Its first applications were the two-loop
calculations of the angular-dependent subset of the electroweak NLLs
for arbitrary processes \cite{Denner:2003wi} and the complete set of
NLLs for the electroweak-singlet massless fermionic form factor
\cite{Pozzorini:2004rm}.  In the future, it can be applied to derive
the two-loop electroweak logarithms for more complicated processes in
order to check the existing resummation prescriptions, which still
rely on arguments based on symmetric gauge theories.  Alternatively,
it can be used to extend the approach that has been used in
\citere{Denner:2001jv} from the one- to the two-loop level, in order
to derive the two-loop NLLs for generic processes within the
spontaneously broken electroweak theory.

\section{Conclusions}

At energies far above the electroweak scale, $\sqrt{s}\gg\MW$, the
electroweak radiative corrections are dominated by large logarithms of
$s/\MW^2$.  These logarithms originate from ultraviolet and mass
singularities.  In this paper, we have presented an algorithm that
permits to extract these singularities and the corresponding large
logarithms from arbitrary multi-loop Feynman integrals by using sector
decomposition.  We have elaborated the algorithm for the case of a
single mass scale and a single energy scale but allowing for various
different mass and energy parameters.  This permits, in particular, to
compute higher-order next-to-leading logarithmic electroweak
corrections for processes involving various kinematical invariants of
the order of hundreds of GeV and masses $\MW\sim\MZ\sim\MH\sim\Mt$ of
the order of the electroweak scale in the approximation where the
masses of the light fermions are neglected.

We have provided explicit formulas for the extraction of the leading
and next-to-leading mass singularities. At the next-to-leading level,
the algorithm has been successfully checked against one-loop and
two-loop results available in the literature, and first applications
have already been published. This method will be useful for the
calculation of next-to-leading logarithmic electroweak two-loop
corrections and for the check of existing resummation prescriptions at
this level.

The method can be extended beyond the next-to-leading level. To this
end, a well-defined set of mass-singular integrals has to be solved in
the required approximation.

\section*{Acknowledgements}

This work was supported in part by the Swiss Bundesamt f\"ur Bildung
und Wissenschaft, by the European Union under contract
HPRN-CT-2000-00149 and by the Deutsche Forschungsgemeinschaft 
in the SFB/TR 09-03.

\section*{Appendix}

\begin{appendix}

\section{Notation and conventions}

For logarithms we introduce the notation
\beqar\label{logsymbdef}
\logarsymbol{\vn}(\ratio):=
\frac{\theta{(\vn)}}{\vn!}
\log^{\vn}(\ratio).
\eeqar
The $\theta$ and the $\intdelta$ functions with integer arguments are
defined as
\beq
\theta{(\vn)} :=
\left\{
\begin{array}{c@{\quad \mbox{for}\quad}\vl} 
1 & \vn \ge 0 \\
0 &
\vn < 0
\end{array}
\right.,
\qquad
\intdelta{(\vn)} :=
\left\{
\begin{array}{c@{\quad \mbox{for}\quad}\vl} 
1 & \vn = 0 \\
0 &
\vn \neq  0
\end{array}
\right. .
\eeq
In order to keep our results as compact as possible 
it is useful to define%
\footnote{Note that $C_\esp=\Psi(\esp)+\gamma_{\mathrm{E}}$, where
  $\Psi(\esp)=\Gamma'(\esp)/\Gamma(\esp)$.  }
\beq\label{Cdefinition}
C_\esp:=
\sum_{\vj=1}^{\esp-1}\frac{1}{\vj} 
.
\eeq
Here and in the following all sums have to be understood as
\beq
\sum_{\vj=\vj_0}^{\vj_1}\equiv \theta(\vj_1-\vj_0)\sum_{\vj=\vj_0}^{\vj_1}
.
\eeq
Similarly, we use the convention $\prod_{\vj=\vj_0}^{\vj_1} \ldots = 1$ for 
$\vj_0>\vj_1$.

For vectors $\vec{\vz}=(\vz_{1},\ldots, \vz_{\vn})$
we introduce the shorthands
\beq\label{productdefinition}
\int_0^1 \mass{\vn}{\vz}{}:=
\int_0^1 \prod_{\vj=1}^{\vn} \rd \vz_{\vj},
\qquad
\product{}{\vz}{}
:=\prod_{\vj=1}^{\vn} \vz_{\vj}
.
\eeq
In the special case $\dim(\vec{\vz})=\vn=0$ the above expressions have
to be understood as
$
\int_0^1 \mass{0}{\vz}{}=
\int_0^1 \prod_{\vj=1}^{0} \rd \vz_{\vj}=1
$ and 
$
\product{}{\vz}{}=
\prod_{\vj=1}^{0} \vz_{\vj}
=1.
$
Powers of vectors $\vec{\vz}=(\vz_1,\dots,\vz_\vn)$ with vector
exponents $\vec{\va}=(\va_1,\ldots,\va_\vn)$ or scalar exponents $\vb$
are defined as
\beq\label{exponentdefinition}
\vec{\vz}^{\,\vec{\va}+\vb}\equiv (\vz_{1}^{\,{\va}_{1}+\vb},\ldots, 
\vz_{\vn}^{\,{\va}_{\vn}+\vb})
.
\eeq 
The ($\vn-1$)-dimensional vector resulting from the omission of the
$\vj$th component of an $\vn$-dimensional vector
$\vec{\vz}=(\vz_1,\dots,\vz_\vn)$ is denoted as
\beq\label{compsuppression}
\vec{\vz}_{[\vj]}=(\vz_1,\dots,\vz_{\vj-1},\vz_{\vj+1},\dots,\vz_\vn).
\eeq
For vectors $\vec{\vr}=(\vr_1,\dots,\vr_\vn)$ with real components and
dimension 
$\vn\ge 0$ 
we introduce the function 
\beqar\label{combinfun}
\combin_\vp(\vec{\vr})&:=&
\theta(\vp) \sum_{\vvb_1,\dots,\vvb_\vn=0}^\vp 
\intdelta \biggl(\vp-\sum_{\vj=1}^\vn \vvb_\vj\biggr) 
\prod_{\vj=1}^\vn \vr_\vj^{\vvb_\vj}
\eeqar
involving all monomials of degree $\vp$.  In the special case
$n=0$ 
we have $\combin_\vp(\vec{\vr})=\intdelta(\vp)$.  Note also that
$\combin_0(\vec{\vr})=1$ for $n\ge 0$.

\section{Massless integrals}
\label{se:onescaleint}
In this section we compute FP integrals of the type
\refeq{genmasslesssdout}, which result from the sector decomposition
of massless loop integrals.  In particular, we perform the singular
integrations over the FPs $\vec{y}$ to NLL accuracy, \ie including the
leading and next-to-leading $1/\varepsilon$ poles.

Let us first consider one-dimensional integrals of the type
\beq
I=\int_0^1 \rd \vy \,\vy^{\vbeta\varepsilon-\vb-1}\vfn(\vy)
\eeq
with $\vbeta$ real and $\vb\ge 0$ integer.  We assume that the
singular behaviour of the integrand at $\vy\to 0$ has been isolated in
the term $\vy^{\vbeta\varepsilon-\vb-1}$ such that $\vfn$ is an
analytic function free of poles and finite at $\vy=0$, \ie
$\vfn(0)\neq 0$.  The $1/\varepsilon$ pole, which originates from an
end-point singularity at $\vy=0$, can be isolated by adding and
subtracting from $\vfn(\vy)$ the term
\beq
\vfnhat(\vy)=\sum_{\vk=0}^{\vb}\vy^\vk\deriv^\vk_\vy\, \vfn(0)
\quad\mbox{with}\quad
\deriv^\vk_\vy \,\vfn(\vy_0)=\left.\frac{1}{\vk!}\frac{\partial^\vk \vfn(\vy)}{\partial \vy^\vk}\right|_{\vy=\vy_0}.
\eeq
To NLL accuracy we obtain
\beqar\label{onedimmasslessres}
I&=&\int_0^1 \rd \vy \,\vy^{\vbeta\varepsilon-\vb-1}\vfnhat(\vy)
+\int_0^1 \rd \vy \,\vy^{\vbeta\varepsilon-\vb-1}\left[\vfn(\vy)-\vfnhat(\vy)\right]
\nl&=&
\sum_{\vk=0}^{\vb}
\frac{1}{\vk-\vb+\beta\varepsilon}\,
\deriv^\vk_\vy \,\vfn(0)
+
\int_0^1 \rd \vy \,\vy^{\vbeta\varepsilon-\vb-1}
\left[\vfn(\vy)-\vfnhat(\vy)\right]
\nl&=&
\frac{1}{\beta\varepsilon}\,
\deriv^\vb_\vy \,\vfn(0)
+
\sum_{\vk=0}^{\vb-1}
\frac{1}{\vk-\vb}\,
\deriv^\vk_\vy\, \vfn(0)
+
\int_0^1 \frac{\rd \vy}{\vy^{\vb+1}}
\left[\vfn(\vy)-\vfnhat(\vy)\right]
+\ord(\varepsilon).
\eeqar
As usual in dimensional regularization, divergent integrals with $b>0$
are defined via analytic continuation.  The generalization to
$\vm$-dimensional integrals is obtained by recursive application of
\refeq{onedimmasslessres}. This yields
\beqar\label{onescaleformulalogsing}
\lefteqn{\int_0^1\mass{\vm}{\vy}{}
\,\product{\vec{\vbeta}\varepsilon-\vec{\vb}-1}{\vy}{}
\vfn(\vec{\vy})}\quad&&
\nl&=&
\left[\product{}{\vbeta}{}\right]^{-1}
\Biggl\{
\frac{1}{\varepsilon^\vm}
{{\deriv^{\vec{\vb}}_{\vec{\vy}}\,\vfn(\vec{0})}}
+\frac{1}{\varepsilon^{\vm-1}}
\sum_{\vi=1}^\vm \vbeta_\vi
\Delta_{\vec\vy}^{\vec\vb,\vi}
\vfn(\vec 0)
+\ord\left(\frac{1}{\varepsilon^{\vm-2}}\right)\Biggr\},
\eeqar
where the derivative operator $\deriv^{\vec{\vb}}_{\vec{\vy}}$ and the
subtraction operator $ \Delta_{\vec\vy}^{\vec\vb,\vi}$ are defined as
\beq\label{multiderivdef}
\deriv^{\vec{\vb}}_{\vec{\vy}}\,
 \vfn(\vec{\vy}_0)=\left.\left[\prod_{\vi=1}^\vm \frac{1}{\vb_\vi!}\right]
\frac{\partial^{\vb_1} }{\partial \vy_1^{\vb_1}}
\dots
\frac{\partial^{\vb_\vm} }{\partial \vy_\vm^{\vb_\vm}}
\vfn(\vec{\vy})
\right|_{\vec{\vy}=\vec{\vy}_0},
\eeq
and
\beqar\label{multimasssubdef}
\lefteqn{
\Delta_{\vec\vy}^{\vec\vb,\vi}\,
\vfn(\vec{\vy}_0)
=
\sum_{\vk_\vi=0}^{\vb_i-1}
\frac{1}{\vk_\vi-\vb_\vi}
\left[\deriv^{\vec{\vb}}_{\vec{\vy}}\,
 \vfn(\vec{y}_0)\right]_{\vb_\vi=\vk_\vi}
}\quad&&\nl&&{}+
\int_0^1 \frac{\rd \vy_\ri}{\vy_\ri^{\vb_\vi+1}}
\Biggl\{
\left[\deriv^{\vec{\vb}}_{\vec{\vy}}\,
\vfn(y_{01},\dots,y_{0\vi-1},\vy_\vi,y_{0\vi+1},\dots,y_{0\vm}) 
\right]_{\vb_\vi= 0}
\nl&&\qquad {}-
\sum_{\vk_\vi=0}^{\vb_i}
\vy_\vi^{\vk_\vi}\left[\deriv^{\vec{\vb}}_{\vec{\vy}}\,
 \vfn(\vec{y}_0)\right]_{\vb_\vi=\vk_\vi}
\Biggr\}.
\nln
\eeqar
The $\vm$ integrations over the FPs $\vec\vy$ in
\refeq{onescaleformulalogsing} yield leading and next-to-leading
singularities of order $1/\varepsilon^\vm$ and
$1/\varepsilon^{\vm-1}$, respectively.  Note that the subtracted terms
\refeq{multimasssubdef}, which enter the coefficients of the
next-to-leading poles, involve one-dimensional integrals over one of
the FPs $\vy_i$.  These integrals are free of the singularity at
$\vy_i=0$ since the corresponding terms within the curly brackets in
\refeq{multimasssubdef} are of order $\vy_i^{\vb_i+1}$.

\section{Massive integrals}
\label{se:twoscaleint}
In this section we compute FP integrals of the type
\refeq{gendecmassiveint1b}, which result
from the sector decomposition of massive loop integrals. In particular,
we focus on the singular integrations over the FPs $\vec{y}$ and
$\vec{z}$.
To be specific, 
we consider the generic integral
\beqar\label{NLLintegraldefd0}
\int_0^1
\mass{\vm}{\vy}{}
\,\product{\vec{\vbeta}\varepsilon-\vec{\vb}-1}{\vy}{}
\int_0^1
\mass{\vn}{\vz}{}
\,\vfn(\vec{\vy};\vec{\vz})\,
\frac{
\product{\vec{\vc}\espt-1+\vec{\vgamma}\varepsilon}{\vz}{}
}{
\left[
\product{\vec{\vc}}{\vz}{}
+\ratio \vfr(\vec{\vy};\vec{\vz})
\right]^{\espt+L\varepsilon}}.
\eeqar
We assume $\vm,\vn$ and $\espt$ integer with $\vm,\vn \ge 0$ and
$\espt>0$ if  $n>0$.%
\footnote{As noted at the end of \refse{se:classif}, 
for $n>0$ we need only the case $\espt>0$.}
 The exponents $\vec\vbeta =(\vbeta_1,\dots,\vbeta_\vm)$ and $\vec\vb
=(\vb_1,\dots,\vb_\vm)$ are integer numbers.  Moreover,
 we assume
$\vb_i\ge 0$ such that the integrations over the FPs
$\vec\vy=(\vy_1,\dots\vy_\vm)$ give rise to a singularity of order
$1/\varepsilon^\vm$.  The exponents
$\vec{\vgamma}=(\vgamma_1,\dots,\vgamma_\vn)$ and $\vec{\vc} \espt
=(\vc_1 \espt,\dots,\vc_\vn \espt)$ are integer numbers with
$\vc_\vj>0$.  The integrations over the FPs
$\vec\vz=(\vz_1,\dots\vz_\vn)$ give rise to mass singularities that
are regulated by the mass-to-energy ratio $w=M^2/s$.

All mass and UV singularities appear as end-point singularities at
$\vec{y}\to \vec{0}$ or $\vec{z}\to \vec{0}$.  As a result of sector
decomposition, all FPs that are responsible for such singularities are
isolated in the terms
$\product{\vec{\vbeta}\varepsilon-\vec{\vb}-1}{\vy}{}$ and
$\product{\vec{\vc}}{\vz}{}$. Thus, $\vfn$ and $\vfr$, which are
analytic functions corresponding to \refeq{h0def} and \refeq{h1def},
are free of end-point singularities and finite at $\vec{y}\to
\vec{0}$, $\vec{z}\to \vec{0}$, \ie $\vfn(\vec{0},\vec{0})\neq 0$ and
$\vfr(\vec{0},\vec{0})\neq 0$.  We also assume that possible poles of
$\vfn$ and $\vfr$ that approach the integration contour as $\ieps\to
0$ can be avoided by appropriate contour deformations.

The integration is performed in the asymptotic limit
\beq
0< \ratio\ll 1
\eeq
to NLL accuracy, as explained in \refse{se:NLLA}.  In the following we
present some intermediate results which correspond to special cases of
\refeq{NLLintegraldefd0} and have been used to compute the above
integral in the most general case:

\begin{itemize}
\item 
Let us start with the special case
$\vfn\equiv\vfr\equiv 1$ and  $\vm=0$.
Here we define 
\beqar\label{basictwoscaledef2}
\Alett{1}{\espt}_{\vn,\vec{\vvb}}(\ratio)
&:=&
\int_0^1
\mass{\vn}{\vz}{}\,
\frac{
\product{\espt-1}{\vz}{}
}{
\left[
\product{}{\vz}{}
+\ratio
\right]^\espt}
 \prod_{\vj=1}^{\vn} \logarsymbol{\vvb_\vj}(\vz_\vj),
\eeqar
where we include combinations of logarithms which originate from the
$\varepsilon$-expansion of the integrand in \refeq{NLLintegraldefd0}.
For $\vec{\vvb}=(\vvb_1,\dots,\vvb_\vn)$ with integer $\vvb_\vj\ge 0$
we obtain
\beqar\label{twoscaleformula12}
\Alett{1}{\espt}_{\vn,\vec{\vvb}}(\ratio)
&\NLL&
(-1)^\vn
\left[
\logarithm{\tilde\vn}{\ratio}
+C_\espt
\logarithm{\tilde\vn-1}{\ratio}
\right]
\quad \mbox{with}\quad \tilde{\vn}=\vn+\sum_{\vj=1}^\vn \vvb_\vj,
\eeqar
where $C_\espt$ is 
defined in \refeq{Cdefinition}.  For $n=0$ the result
\refeq{twoscaleformula12} is trivial, since
\beqar
\Alett{1}{\espt}_{0}(\ratio)
&=&\frac{1}{(1+\ratio)^\espt}
=1+\ord(\ratio).
\eeqar
For $n\ge 1$ and  $\espt=1$ we have a polylogarithm
\beqar
\Alett{1}{1}_{n,\vec \vvb}(\ratio)
&=&
(-1)^{\tilde{n}-n+1}\,
\mathrm{Li}_{\tilde\vn}\left(-\frac{1}{\ratio}\right)
\nl&=&
(-1)^{n}\,
\logarithm{\tilde\vn}{\ratio}
+\ord\left[
\logarithm{\tilde\vn-2}{\ratio}
\right]
+\ord\left(
\ratio
\right).
\eeqar
For $n\ge 1$ and $\espt\ge 2$, \refeq{twoscaleformula12}
can be proven by induction using the recursive relation
\beqar
\Alett{1}{\espt}_{\vn,(\vvb_1,\dots,\vvb_\vn)}
(\ratio)
&=&
\Alett{1}{\espt-1}_{\vn,(\vvb_1,\dots,\vvb_\vn)}(\ratio)
+\frac{1}{\espt-1}\Biggl[
\Alett{1}{\espt-1}_{\vn,(\vvb_1,\dots,\vvb_{\vn-1},\vvb_\vn-1)}(\ratio)
\nl&&{}
-\intdelta(\vvb_\vn)
\Alett{1}{\espt-1}_{\vn-1,(\vvb_1,\dots,\vvb_{\vn-1})}(\ratio)
\Biggr],
\eeqar
which follows from partial integration in $\vz_\vn$.
\item 
Using \refeq{twoscaleformula12} one can easily prove that 
\beqar\label{application1}
\int_0^1
\mass{\vn}{\vz}{}
\frac{
\product{\espt-1}{\vz}{}
}{
\left(
\product{}{\vz}{}
+\ratio\right)^\espt}
\;
\logarsymbol{\vp}\left(
\product{\vec{\vr}}{\vz}{}
\right)
\NLL
(-1)^\vn\combin_\vp(\vec{\vr})
\left[
\logarithm{\vn+\vp}{\ratio}
+C_\espt\logarithm{\vn+\vp-1}{\ratio}
\right]
\nln
\eeqar
for any real vector $\vec{\vr}=(\vr_1,\dots,\vr_\vn)$ and integer
number $\vp$, where $\combin_\vp(\vec{\vr})$ is the combinatorial
factor defined in \refeq{combinfun}.

\item 
In the special case $\vfn \equiv \vfr \equiv 1$ and $\vec\vb =\vec 0$ we obtain
\beqar\label{twoscaleformula12d}
\lefteqn{
\int_0^1
\mass{\vm}{\vy}{}
\,\product{\vec{\vbeta}\varepsilon-1}{\vy}{}
\int_0^1
\mass{\vn}{\vz}{}
\frac{
\product{\vec{\vc}\espt-1+\vec{\vgamma}\varepsilon}{\vz}{}
}{
\left[
\product{\vec{\vc}}{\vz}{}
+\ratio \right]^{\espt+L\varepsilon}}
}\quad&&\nl
&\NLL&{}
\left[\product{}{\vbeta}{}\product{}{\vc}{}\right]^{-1}
(-1)^\vn \sum_{\vp\ge 0}
\combin_\vp(\vec{\vr})
\left(\frac{1}{\varepsilon}\right)^{\vm-\vp}
\left[
\logarithm{\vn+\vp}{\ratio}
+C_\espt\logarithm{\vn+\vp-1}{\ratio}
\right],
\nln
\eeqar
where the components of the vector  $\vec{\vr}=(\vr_1,\dots,\vr_\vn)$ read
\beq\label{rvector}
\vr_\vj=\frac{\vgamma_\vj}{\vc_\vj}-L.
\eeq
The integration of the variables $\vy_1,\dots,\vy_\vm$ in
\refeq{twoscaleformula12d} is trivial and gives rise to a pole of
order $1/\varepsilon^\vm$.  The remaining integrand has to be expanded
up to the needed order in $\varepsilon$.  Then, after neglecting
irrelevant terms of order $\ratio$, it can be integrated with the help
of \refeq{application1}.

\item In the special case $m=0$, including logarithms as in
  \refeq{basictwoscaledef2}, we obtain
\beqar\label{NLLscaleresult}
&&\int_0^1
\mass{\vn}{\vz}{}
\,\vfn(\vec{\vz})\,
\frac{
\product{\espt-1}{\vz}{}
}{
\biggl[
\product{}{\vz}{}
+\ratio \vfr(\vec{\vz})\biggr]^\espt}
 \prod_{\vj=1}^{\vn} \logarsymbol{\vvb_\vj}(\vz_\vj)
\NLL
(-1)^\vn 
\Biggl\{
{\vfn(\vec{0})}
\logarithm{\tilde{\vn}}
{\ratio \vfr(\vec{0})}
\nl
&&\quad{}+
\left[
{\vfn(\vec{0})}C_\espt
-
\sum_{\vj=1}^{\vn}\intdelta(\vvb_\vj)
\Delta_{\vec\vz}^{\vec 0,\vj}\vfn(\vec{0})
\right]
\logarithm{\tilde{\vn}-1}{\ratio}
\Biggr\},
\eeqar
where  $\tilde{\vn}=\vn+\sum_{\vj=1}^\vn \vvb_\vj$
and analogously to  \refeq{multimasssubdef} the subtraction operator $\Delta_{\vec \vz}^{\vec 0,\vj}$
is defined as
\beq\label{multimasssubdef2}
\Delta_{\vec \vz}^{\vec 0,\vj}\vfn(\vec \vz_0)
=
\int_0^1\frac{\rd \vz_{\vj}}{\vz_{\vj}}
\left[
\vfn(\vz_{01},\dots,\vz_{0\vj-1},\, \vz_{\vj},\vz_{0\vj+1},\dots,\vz_{0\vn})
-
\vfn(\vec \vz_{0})
\right].
\eeq

The above result has been obtained by means of the subtraction
technique discussed in \refse{se:subtraction}.  Each integration has
been split into singular and non-singular parts as
\beqar\label{intexpa}
\lefteqn{\int_0^1 \rd \vz_\vj I(\vfn,\vfr;\vec{\vz})
=
\left.\int_0^1 \rd \vz_\vj I(\vfn,\vfr;\vec{\vz})\right|_{h_\vi\equiv h_\vi(\vz_1,\dots,\vz_{\vj-1},0,\vz_{\vj+1},\dots,\vz_\vn)}
}\quad&&
\nl&&{}+
\int_0^1 \rd \vz_\vj \left[\left.
I(\vfn,\vfr;\vec{\vz})-I(\vfn,\vfr;\vec{\vz})\right|_{h_\vi\equiv h_\vi(\vz_1,\dots,\vz_{\vj-1},0,\vz_{\vj+1},\dots,\vz_\vn)}\right],
\eeqar
where $I(\vfn,\vfr;\vec{\vz})$ represents the integrand on the
left-hand side of \refeq{NLLscaleresult}.  The recursive application
of this subtraction to all $\vn$ integrations gives rise to various
terms, which consist of combinations of $\vl$ singular parts and
$\vn-\vl$ non-singular parts
with $0\le \vl\le \vn$.  The LL and NLL 
contributions correspond to the terms with 
$\vl=\vn,\vn-1$,
and read
\beqar\label{intexp}
\lefteqn{\int_0^1
\mass{\vn}{\vz}{}\,
I(\vfn,\vfr;\vec{\vz})
 \NLL \left.\int_0^1
\mass{\vn}{\vz}{}\,
I(\vfn,\vfr;\vec{\vz})
\right|_{h_\vi\equiv h_\vi(\vec{0})}
}\quad&&\nl
&&{}+\sum_{\vj=1}^{\vn}\int_0^1
\mass{\vn}{\vz}{}
\left[
\left.I(\vfn,\vfr;\vec{\vz})
\right|_{h_\vi\equiv h_\vi(0,\dots,0,\, \vz_{\vj},0,\dots,0)}-
\left.I(\vfn,\vfr;\vec{\vz})
\right|_{h_\vi\equiv h_\vi(\vec{0})}\right].
\eeqar
The first term on the right-hand-side of \refeq{NLLscaleresult}
originates from the first term on the right-hand-side of
\refeq{intexp} and has to be expanded to NLL accuracy as
\beqar
\logarithm{\tilde{\vn}}{\ratio \vfr(\vec{0})}
\NLL
\logarithm{\tilde{\vn}}{\ratio}
+
\log{\left({\vfr(\vec{0})}\right)}
\logarithm{\tilde{\vn}-1}{\ratio}.
\eeqar
The second term on the right-hand-side of \refeq{intexp} gives rise to
the $\Delta$ contributions in \refeq{NLLscaleresult}.  Here the
function $\intdelta(\vvb_\vj)$ indicates that only terms with
$\vvb_\vj=0$ yield a NLL contribution.

\item Using \refeq{NLLscaleresult} it is easy to prove that for any
  real vector $\vec{\vr}=(\vr_1,\dots,\vr_\vn)$ and integer number
  $\vp$,
\beqar\label{NLLintegraldefappli}
&&\int_0^1
\mass{\vn}{\vz}{}
\,\vfn(\vec{\vz})\,
\frac{
\product{\espt-1}{\vz}{}
}{
\biggl[
\product{}{\vz}{}
+\ratio \vfr(\vec{\vz})\biggr]^\espt}
\logarsymbol{\vp}\left(
\product{\vec{\vr}}{\vz}{}
\right)\NLL
(-1)^\vn\Biggl\{
{\vfn(\vec{0})}
\combin_\vp(\vec{\vr})
\logarsymbol{\vn+\vp}
\left({\ratio}\vfr(\vec{0})\right)
\nl
&&\hspace{10mm}
{}+\left[
{\vfn(\vec{0})}
\combin_\vp(\vec{\vr})C_\espt
-
\sum_{\vj=1}^{\vn}
\combin_\vp(\vec{\vr}_{[\vj]})
\Delta_{\vec\vz}^{\vec 0,\vj}{\vfn(\vec{0})}
\right]\logarsymbol{\vn+\vp-1}
\left(\ratio\right)
\Biggr\},
\eeqar
where $\vec{\vr}_{[\vj]}$ 
and $\combin_\vp$ are defined in \refeq{compsuppression} and
\refeq{combinfun}, respectively.
\end{itemize}
Finally, combining \refeq{onescaleformulalogsing} and
\refeq{NLLintegraldefappli} we arrive at the result
\beqar\label{NLLscaleresultd}
\lefteqn{ \int_0^1
\mass{\vm}{\vy}{}
\,\product{\vec{\vbeta}\varepsilon-\vec{\vb}-1}{\vy}{}
\int_0^1
\mass{\vn}{\vz}{}
\,\vfn(\vec{\vy};\vec{\vz})\,
\frac{
\product{\vec{\vc}\espt-1+\vec{\vgamma}\varepsilon}{\vz}{}
}{
\left[
\product{\vec{\vc}}{\vz}{}
+\ratio \vfr(\vec{\vy};\vec{\vz})
\right]^{\espt+L\varepsilon}}
=}\quad&&\nl
&\NLL&
\left[\product{}{\vbeta}{} \product{}{\vc}{}\right]^{-1}
(-1)^\vn\sum_{\vp\ge0}
\left(\frac{1}{\varepsilon}\right)^{\vm-\vp}
\Biggl\{
\combin_\vp(\vec{\vr})
\deriv^{\vec{\vb}}_{\vec{\vy}}\,
{\vfn^{(0)}(\vec{0};\vec{0})}
\logarsymbol {\vn+\vp}\left(\ratio\right)
\nl&&{}
+ \Biggl[
\theta(\vn-1)\combin_\vp(\vec{\vr})
\deriv^{\vec{\vb}}_{\vec{\vy}}
\left[
{\vfn^{(0)}(\vec{0};\vec{0})}
\left[
\log{\left({\vfr(\vec{0};\vec{0})}\right)}
+C_\espt\right]\right]
\nl&&{}-
\sum_{\vj=1}^\vn \vc_\vj
\combin_{\vp}(\vec{\vr}_{[\vj]}) 
\Delta_{\vec\vz}^{\vec 0,\vj}
\deriv^{\vec{\vb}}_{\vec{\vy}}\,
{\vfn^{(0)}(\vec{0};\vec{0})}
\nl&&{}+\combin_{\vp-1}(\vec{\vr})
\left[
\deriv^{\vec{\vb}}_{\vec{\vy}}\,
{\vfn^{(1)}(\vec{0};\vec{0})}
+\sum_{\vi=1}^\vm \vbeta _\vi \Delta_{\vec\vy}^{\vec\vb,\vi}
{\vfn^{(0)}(\vec{0};\vec{0})}\right]\Biggr]
\logarsymbol{\vn+\vp-1}\left(\ratio\right)
\Biggr\}
,
\eeqar
where
 \beq
\vfn
\equiv
\sum_{\vj=0}^\infty \frac{\varepsilon^\vj}{\vj!}\vfn^{(\vj)}
\eeq
is the $\varepsilon$-expansion%
\footnote{The leading term $\vfn^{(0)}(\vec{\vy};\vec{\vz})$ of this
  $\varepsilon$ expansion can be of order $1/\varepsilon^s$.}  of the
function $\vfn\equiv\vfn(\vec{\vy};\vec{\vz})$.  The vector
$\vec{\vr}=(\vr_1\dots,\vr_\vn)$, $\vec{\vr}_{[\vj]}$, and
$\combin_\vp$ are defined in \refeq{rvector}, \refeq{compsuppression},
and \refeq{combinfun}, respectively. Moreover, the derivative operator
$\deriv^{\vec{\vb}}_{\vec{\vy}}$ and the subtraction operators $
\Delta_{\vec\vy}^{\vec\vb,\vi}$ and $\Delta_{\vec\vz}^{\vec 0,\vj}$
are defined in \refeq{multiderivdef}, \refeq{multimasssubdef} and
\refeq{multimasssubdef2}, respectively.  In particular, we have
\beqar
\Delta_{\vec \vz}^{\vec0,\vj}
\deriv^{\vec{\vb}}_{\vec{\vy}}\,
{\vfn^{(0)}(\vec{0};\vec{0})}
&=&\int_0^1\frac{\rd \vz_{\vj}}{\vz_{\vj}}
\left[
\deriv^{\vec{\vb}}_{\vec{\vy}}\,
\vfn^{(0)}(\vec{0};0,\dots, \vz_{\vj},\dots,0)
-
\deriv^{\vec{\vb}}_{\vec{\vy}}\,
{\vfn^{(0)}(\vec{0};\vec{0})}
\right].
\eeqar
The $\vm+\vn$ integrations over the FPs $\vec\vy$ and $\vec\vz$ in
\refeq{NLLscaleresultd} yield leading and next-to-leading
singularities of order $\varepsilon^{-\vm+p} \log^{\vn+p}(\ratio)$ and
$\varepsilon^{-\vm+p} \log^{\vn+p-1}(\ratio)$, respectively.
Similarly to the massless case, the subtracted $\Delta$ terms which
enter the coefficients of the next-to-leading poles, involve
convergent one-dimensional integrals over one of the FPs $\vy_i$ or
$\vz_j$.

We note that in the trivial case $n=0$ the integral
\refeq{NLLintegraldefd0} simplifies to
\beqar\label{NLLintegraldefd0massless}
\int_0^1
\mass{\vm}{\vy}{}
\,\product{\vec{\vbeta}\varepsilon-\vec{\vb}-1}{\vy}{}
\frac{\vfn(\vec{\vy})}{
\left[1+\ratio \vfr(\vec{\vy})\right]^{\espt+L\varepsilon}}=
\int_0^1
\mass{\vm}{\vy}{}
\,\product{\vec{\vbeta}\varepsilon-\vec{\vb}-1}{\vy}{}
{\vfn(\vec{\vy})}
+\ord(\ratio).
\eeqar
Since no singularity is regulated by the mass term $\ratio$, in the
asymptotic limit $\ratio\to 0$ \refeq{NLLintegraldefd0massless}
corresponds to the massless integral that is obtained by setting
$\ratio=0$ in the integrand.  Indeed, in the special case $n=0$, the
general result \refeq{NLLscaleresultd} for massive integrals is
equivalent to the result \refeq{onescaleformulalogsing} for massless
integrals, as one can easily verify using
$\combin_\vp(\vec{\vr})=\intdelta(\vp)$ for $\dim{(\vec{\vr})}=n=0$.

\section{Factorization of $1/\varepsilon$ poles from massive
  integrals}
\label{se:ibp}

In this section, we consider again the generic sector integrals
\refeq{cstdsdout1}--\refeq{factparexp}, which have been computed in
\refse{se:massivedecomp} assuming \refeq{Tcase}, such that all
singularities resulting from the integrations over the FPs
$\xi_1,\dots,\xi_p$, \ie the FPs that have been factorized in
\refeq{decden}, are regulated by the mass term $\ratio$ in
\refeq{decden}.  Here we discuss the more general case where
$\vT_\vk\le -1$ for some $1\le k \le \vp$ and the corresponding
$\xi_k$-integrations yield additional $1/\varepsilon$ poles.  As we
have already anticipated, this kind of integrals can be eliminated by
means of (recursive) integration by parts in the variables $\xi_\vk$,
which permits to increase the exponents $\vT_\vk\le -1$ until the
$1/\varepsilon$ poles are factorized and only integrals respecting
\refeq{Tcase} remain.

In the following we consider sector integrals
\refeq{cstdsdout1}--\refeq{factparexp} with \refeq{dcase} assuming
that possible non-logarithmic singularities have been eliminated as
described in \refse{se:remapping}.  An integration by parts in the
variable $\xi_k$ yields
\beqar\label{intbyparts}
\vGt(\vec\vT,\espt,\vgt,\Ft)
&=&
\frac{1}{\vT_k+1+\tau_k\varepsilon}\int_0^1\mass{I-1}{\xi}{}
\,\product{\vec{\vT}+\vec{\vtau}\varepsilon}{\xi}{}
\left[\delta(1-\xi_k)
-\xi_k {\partial_{\xi_k}}
\right]
\left(\frac{\vgt}{\Ft^{\espt+L\varepsilon}}\right)
,\quad
\eeqar
where the $\vec\xi$-dependence of $\vgt$ and $\Ft$ is implicitly understood.
The $\delta(1-\xi_k)$
contribution on the right-hand side represents the boundary 
term at $\xi_k=1$, whereas the boundary term at $\xi_k=0$ vanishes.%
\footnote{This holds for arbitrary values of $\vT_k$ within
  dimensional regularization, \ie also for $\vT_k< -1$ where the
  boundary term at $\xi_k=0$ is divergent for $\varepsilon=0$.}  The
remaining $\xi_k\partial_{\xi_k}$
contribution being proportional to
$\xi_k$ has the desired property to increase the power $T_k$
associated with the FP $\xi_k$.  However, as one can see from
\beqar\label{derivterm}
\xi_k{\partial_{\xi_k}} 
\left(\frac{\vgt}{\Ft^{\espt+L\varepsilon}}\right)=
-
\frac{({\espt+L\varepsilon})\vgt}
{\Ft^{\espt+1+L\varepsilon}}
\left\{
\product{\vec{\vt}}{\xi}{}
\left[{\vt_k}+\xi_k\partial_{\xi_k}\right]\vfst+
\ratio \xi_k \partial_{\xi_k}\vfmt
\right\}
+
\frac{\xi_k{\partial_{\xi_k}}  \vgt}
{\Ft^{\espt+L\varepsilon}},
\eeqar
the degrees of singularity $d_j=\espt-(\vT_j+1)/\vt_j$  
of the various FP integrations $1\le j \le \vp$ can increase for the
term proportional to $\ratio \xi_k{\partial_{\xi_k}}\vfmt$,
where the increase of $\espt$ is not compensated by the factor
$\product{\vec{\vt}}{\xi}{}$ in the numerator.  In order to preserve
\refeq{dcase}, this must be avoided by rewriting
\beqar\label{derivmassterm}
\ratio \xi_k{\partial_{\xi_k}}\vfmt =
\frac{\xi_k(\partial_{\xi_k}\vfmt)}{\vfmt-\ieps} \left[\Ft-
  \product{\vec{\vt}}{\xi}{}\vfst\right],
\eeqar
in \refeq{derivterm}.  The integration by parts in the FP $\xi_k$
combined with \refeq{derivmassterm} results in a sum of three
contributions,
\beqar\label{lincombibp}
\vGt(\vec\vT,\espt,\vgt,\Ft)=\frac{1}{\vT_k+1+\tau_k\varepsilon}\sum_{\vr=1}^3 \,\vGt(\vec\vT_\vr,\espt_\vr,\vgt_\vr,\Ft),
\eeqar
that have the same structure as the original integral with
\beqar\label{ibpsubst}
\espt_1&=&\espt,\qquad
\vec\vT_1=\vec\vT 
,\qquad
\vgt_1={ \delta(1-\xi_k)\,\vgt},\nl
\espt_2&=&\espt+1,\qquad
\vec\vT_2=
\vec\vT+\vec\vt,\qquad
\vgt_2={(\espt +L\varepsilon)\,\vgt}
\left[{\vt_k}-\frac{\xi_k (\partial_{\xi_k} \vfmt)}{\vfmt-\ieps}
+\xi_k\partial_{\xi_k}\right]\vfst,\nl
\espt_3&=&\espt,\qquad
\left(\vec\vT_3\right)_j=
\vT_j+\delta_{jk}
,\qquad
\vgt_3={(\espt +L\varepsilon)}
\frac{(\partial_{\xi_k} \vfmt)}{\vfmt-\ieps} \vgt
-\partial_{\xi_k}\vgt.
\eeqar
The exponent $\vT_k$ associated with the FP $\xi_k$ grows in all
terms, \ie $(\vec \vT_\vr)_k>\vT_k$, apart from the boundary term
$\vGt(\vec\vT_1,\espt_1,\vgt_1,\Ft)$, where the $\xi_k$ integration is
irrelevant.  Thus, in order to eliminate all integrals that do not
satisfy \refeq{Tcase}, it is in principle sufficient to iterate the
integration by parts to all FPs $\xi_k$ with $\vT_k\le -1$.  However,
since our aim is to reduce by one the number of integrations every
time that we extract a singularity, care must be taken when
$1/\varepsilon$ poles are extracted through the overall factor
$1/(\vT_k+1+\vtau\varepsilon)$ in \refeq{lincombibp}, \ie every time
that we perform an integration by parts in a FP $\xi_k$ with
$\vT_k=-1$.  In practice, as we show below, it is possible to restrict
the use of \refeq{lincombibp}--\refeq{ibpsubst} to the cases where
$\espt=0$ and $\partial_{\xi_k}\vgt=0$ if $\vT_k=-1$, such that
$\vGt(\vec\vT_\vr,\espt_\vr,\vgt_\vr,\Ft)=\ord(\varepsilon)$ for
$\vr=2,3$ and the only $1/\varepsilon$ singularity results from the
boundary term $\vGt(\vec\vT_1,\espt_1,\vgt_1,\Ft)$.

Let us now outline our general strategy to eliminate all contributions
that do not satisfy \refeq{dcase} and/or \refeq{Tcase}.  Starting from
a generic sector integral \refeq{cstdsdout1}--\refeq{factparexp} we
proceed as follows:
\begin{enumerate}
\item If non-logarithmic singularities are present, \ie $d>0$, 
we eliminate them  by means of \refeq{lincomb}--\refeq{nonlogtransf},
such that the resulting integrals satisfy \refeq{dcase}.

\item If there is a FP $\xi_k$ with $1\le k\le \vp$ and $\vT_k<-1$,
  this FP is integrated by parts using
  \refeq{lincombibp}--\refeq{ibpsubst}. This step is repeated until all
  integrals satisfy $\vT_k\ge -1$ for $1\le k \le \vp$. Here no
  $1/\varepsilon$ pole appears.
  
\item For the remaining integrals involving a $\xi_k$ with $1\le k\le
  \vp$ and $\vT_k=-1$ we have $\espt=d_k \le d \le 0$ as a result of
  step 1 and owing to \refeq{singdegdef}.
If $\espt<0$, then 
\beq
\Ft^{-(\espt+L\varepsilon)}=
\Ft^{-L\varepsilon}\left\{\left[\product{\vec{\vt}}{\xi}{}\vfst\right]^{-\espt}
+\ord(\ratio)\right\},
\eeq
and the contributions of order $\ratio$ can be omitted. Thus, we can
rewrite the corresponding integral as
\beqar
\vGt(\vec\vT,\espt,\vgt,\Ft)=\vGt(\vec\vT-\espt\vec\vt,0,\vgt \vfst^{-\espt},\Ft),
\eeqar
and the integral on the right-hand side, with exponents
$\vec\vT\to \vec\vT-\espt\vec\vt$, already satisfies \refeq{Tcase} as
a consequence of step 2 ($\vT_k\ge -1$) and $\espt<0$.

\item Finally, we have to consider the integrals with $\espt=0$ and
  a $\xi_k$ with $1\le k\le \vp$ and $\vT_k=-1$.  In this
  case, we split $\vgt$ as
\beqar
\vgt(\vec{\xi})=\vgt_k(\vec{\xi})
+\Delta \vgt_k(\vec{\xi})
\quad\mbox{with}\quad
\vgt_k(\vec{\xi})=\vgt(\xi_1,\dots,\xi_{k-1},0,\xi_{k+1},\dots,\xi_{I-1})
\eeqar
and treat these two contributions separately, \ie
\beqar
\vGt(\vec\vT,\espt,\vgt,\Ft)=
\vGt(\vec\vT,\espt,\vgt_k,\Ft)+
\vGt(\vec\vT,\espt,\Delta \vgt_k,\Ft).
\eeqar
On the one hand, for the contribution $\vGt(\vec\vT,\espt,\vgt_k,\Ft)$
we can safely perform an integration by parts in $\xi_k$ since, as
discussed above, owing to $\partial_{\xi_k}\vgt_k(\vec{\xi})=0$ and
$\espt=0$, only the boundary term effectively contributes to the
resulting $1/\varepsilon$ singularity.  On the other hand, the
remaining contribution is free of $\xi_k$~singularities owing to
\beqar
\lim_{\xi_k\to 0} \xi_k^{-1}\Delta\vgt_k(\vec{\xi})&=&
\left.\partial_{\xi_k}\vgt(\vec{\xi})\right|_{\xi_k=0},
\eeqar
and can thus be written as 
\beq
\vGt(\vec\vT,\espt,\Delta \vgt_k,\Ft)=
\vGt(\vec\vT',\espt,\xi_k^{-1}\Delta \vgt_k,\Ft),
\eeq
where $\vec\vT'=(\vT_1,\dots,\vT_{k-1},0,\vT_{k+1},\dots,\vT_{I-1})$.
The power $\vT_k$ of $\xi_k$ has been reduced in both contributions
without generating unwanted $1/\varepsilon$ poles.  At the end, if
other FPs $\xi_j$ with $1\le j\le \vp$ and $\vT_j=-1$ are still
present this step has to be iterated until all integrals fulfil
\refeq{Tcase}.
\end{enumerate}

\end{appendix}

\end{document}